\documentclass[12pt]{elsarticle}
\usepackage[utf8]{inputenc} 
\usepackage{amssymb}
\usepackage{amsmath,color}
\usepackage{algorithm}
\usepackage{algpseudocode}
\usepackage{lineno}
\usepackage{subcaption}

\newcommand{\R}{\mathbb{R}}

\newcommand{\pp}[2]{\frac{\partial {#1}}{\partial {#2}}}
\newcommand{\pptwo}[2]{\frac{\partial^2 {#1}}{\partial {#2}^2}}

\renewcommand{\vec}[1]{\mathbf{#1}}
\newcommand{\vecg}[1]{\boldsymbol{#1}}

\newcommand{\Sdp}{\widehat{\boldsymbol{\mathcal{S}}}^{DP}_{\phi_{\epsilon}}}

\newcommand{\Stokes}{\boldsymbol{\mathcal{S}}_{\phi_{\epsilon}}} 

\newcommand{\Zmodn}{\mathbb{Z}_N} 
\newcommand{\Fmodn}{\mathbb{F}_N}

\newcommand{\Kd}{K^D_{\alpha,\beta}}
\newcommand{\whvec}[1]{\widehat{\vec{#1}}}

\newcommand{\tilalpha}{\tilde{\alpha}}
\newcommand{\tilbeta}{\tilde{\beta}}

\journal{Journal of Computational Physics}
\begin{document}
\makeatother

\begin{frontmatter}
\title{Analysis of the stability of an immersed elastic surface using the method of regularized Stokeslets}
\author{Dana Ferranti and Sarah D. Olson}
\affiliation{organization={Department of Mathematical Sciences, Worcester Polytechnic Institute},addressline={100 Institute Rd}, 
            city={Worcester},
            postcode={01609}, 
            state={MA},
            country={USA}}

            \begin{abstract} 
            A linear stability analysis of an elastic surface immersed in a viscous fluid is presented. The coupled system is modeled using the method of regularized Stokeslets (MRS), a Lagrangian method for simulating fluid-structure interaction at zero Reynolds number. The linearized system is solved in a doubly periodic domain in a 3D fluid. The eigenvalues determine the theoretical critical time step for numerical stability for a forward Euler time integration, which are then verified numerically across several regularization functions, elastic models, and parameter choices. New doubly periodic regularized Stokeslets are presented, allowing for comparison of the stability properties of different regularization functions. The stability results for a common regularization function are approximated by a power law relating the regularization parameter and the surface discretization for two different elastic models. This relationship is empirically shown to hold in the different setting of a finite surface in a bulk fluid.
            \end{abstract}



\end{frontmatter}

\section{Introduction}
Dynamic and elastic structures that are immersed in a fluid are a common type of fluid-structure interaction problem \cite{Griffith20}. Spermatozoa (sperm) and bacteria \cite{gaffney11,Lauga16}, cilia beating within airways \cite{Gilpin20,Salathe07}, and microtubules in the cytoplasm during cell division  and cell migration \cite{Forth17,Kent17} are examples of elastic biological structures at the micro-scale. Cells such as platelets and red blood cells in the blood are also surrounded by an elastic membrane or surface that is deformable \cite{Griffith20,Freund14}. In addition, artificial capsules and microrobots are being developed for drug delivery \cite{Biesel16,Nelson23}. 
These structures may deform in response to fluid flows or other forces (e.g. mechanical or chemical) and may interact with other surfaces or elastic structures \cite{Espina23,Rallabandi24}. 

To address questions related to the interaction of these elastic structures with the surrounding fluid, modeling approaches have to account for accurate geometries and deformations that are fully coupled to the fluid. Methods utilizing boundary integral equations \cite{pozr1992, veerapaneni09}, the immersed boundary method \cite{peskin}, and the method of regularized Stokeslets (MRS) \cite{cortez2001,cortez2005} are commonly used to model an elastic structure immersed in a viscous, incompressible fluid.  The classic boundary integral formulation reduces the computation of a 3-dimensional (3D) flow to evaluation of a surface integral, but when computing velocities very near a boundary, one most evaluate integrals with singular kernels. The IB method requires both a Lagrangian grid (moves with the elastic body) and Eulerian grid (fixed) for the unknowns and computation of velocities on a 3D fixed grid can be computationally expensive. The MRS is in between these methods; it uses a Lagrangian representation of the structure but does not require an Eulerian grid for the fluid, and it can be cast in the form of a boundary integral equation \cite{cortez2005, smith18, smithDouble}.

For structures at the micro-scale, we are in the regime where viscous forces dominate inertial forces \cite{gaffney11,Needleman19} and  
we will assume the fluid is governed by the incompressible Stokes equations, 
\begin{subequations}
    \begin{align}
        \vec 0 &= -\nabla p (\vec x) + \mu \Delta \vec u(\vec x) + \vec f(\vec x) \label{eq:st}\\ 
        0 &=\nabla \cdot \vec u(\vec x),\label{eq:div}
    \end{align}
    \label{eq:stokes}
\end{subequations}
\hspace{-.32cm} where $\vec u = (u, v, w)$ is the fluid velocity, $p$ is the fluid pressure, $\mu$ is the dynamic viscosity,  $\vec f(\vec x)$ is a force per unit volume, 
and $\vec x\in\mathbb{R}^3$. 

We use a Lagrangian description for an elastic body immersed in the fluid, $\vec X(\vec q,t)$, where $\vec q$ are material coordinates in a reference configuration and $t$ is time. In the case that $\vec X$ is a 2D surface, $\vec q=(q,s)$, and if $\vec X$ is a 1D space curve, $\vec q=q$, a scalar. The internal forces exerted by the structure on the fluid are represented as a force density, $\vec{F}(\vec X(\vec q,t))$. We will assume the structure is neutrally buoyant and satisfies the no-slip condition

\begin{equation}
\label{eq:noslip}
    \vec u(\vec X)=\frac{\partial \vec X}{\partial t},
\end{equation}
enforcing the structure velocity matches the local fluid velocity at their interface.

A significant computational challenge for the simulation of elastic structures in viscous fluids is the numerical stiffness that arises due to the structures' tensile and bending rigidity (felt in the fluid equation, \eqref{eq:st}, via $\vec f$). For a discretized structure, the no-slip condition in \eqref{eq:noslip} 
leads to a system of stiff initial value problems to determine the new shape of the elastic structure. Hence, this numerical stiffness is a common feature for numerical implementations of slender body theory \cite{tornberg2004, nazockdast2017}, the immersed boundary method \cite{huapeskin,Gong08,stockie99}, and the MRS. As a result, severe time-step restrictions exist for explicit methods \cite{stockie95,stockie99,huapeskin}. On the other hand, implicit methods correspond to a nonlinear system that is computationally intensive to solve \cite{Guy15,nazockdast2017}. 

In this study, we focus on the MRS. It is a popular method due to its ease of implementation and extensions to
several fluid domains via the utilization of different regularized fundamental solutions \cite{ainley,hoffmann2,cortezvarela,hoffmann1,LBCL2013,mannan,tripBrinkman,Zheng23}. There are several ways to derive regularized Stokeslets, but at the heart of the method, it requires the utilization of a radially symmetric mollifier or blob function $\phi_{\epsilon}$ that spreads the singular force layer from the elastic structure to the fluid. A common example of an algebraic blob function in 3D satisfying $\iiint_{\mathbb{R}^3}\phi_\epsilon(\vec x,\vec X) d\mathbf{x}=1$ is \begin{equation}
 \phi_{\epsilon}(r)=\frac{15\epsilon^4}{8\pi(r^2+\epsilon^2)^{7/2}} \label{eq:algblob}
\end{equation}
for $r=|\vec x-\vec X|$ (standard Euclidean norm) and regularization parameter $\epsilon$ that controls the width over where the majority of the force is spread \cite{cortez2005}. 
By regularizing the forces in \eqref{eq:st} or directly regularizing the fundamental solution, the Stokeslet,
we can determine an exact solution for a given point force 
to solve for the resulting fluid velocity at any point in the fluid domain or on the structure. 

Within the framework of the MRS, approaches that have been taken to alleviate the numerical stiffness and time step restrictions include explicit multirate time integration with spectral deferred corrections \cite{bouzarth2010}, parallel-in-time methods \cite{Liu22}, multigrid methods \cite{Liu23}, and in the case of Kirchoff rods, removing the Lagrange multiplier formulation of the forces to maintain inextensibility \cite{jabbarzadeh2020}. However, we note that appropriately handling the time-step restriction has to be completed in conjunction with minimizing other types of error which depend on the choice of blob function and the size of the regularization parameter \cite {cortez2005, nguyen2014, zhao2019, Chisholm22}, the discretization of the structure \cite{nguyen2025}, and the method used to integrate the regularized Stokeslets over the structure \cite{bgil, smith18, cortez18, ferranti2024}.

While it is well known that there are severe time step restrictions with explicit methods using the MRS, this is a widely used method with no guidance on time step choice for the practitioner. To date, we are not aware of any numerical stability analysis for the MRS and, in particular, how stability relates to choices in the elastic force model (tension and/or bending), the regularization function $\phi_{\epsilon}$, and other model/numerical parameters. This is a first attempt at such a study.

 We will focus on a linear stability analysis of a representative structure, an elastic surface with a small initial perturbation from its flat resting configuration that is immersed in a fluid governed by the incompressible Stokes equations given in \eqref{eq:stokes}. The analysis and problem setup is inspired by previous work in the immersed boundary (IB) literature where the setting was a 1D elastic fiber in a 
2D fluid or a 2D elastic surface in a 3D fluid \cite{Boffi07,Heltai08,huapeskin,stockie95,stockie99,Gong08}. 
 A linearization of the fluid equations and boundary conditions was utilized to enable the analysis. 
 Earlier studies  based on Fourier analysis did not focus on boundary discretization \cite{stockie95,stockie99, Gong08} whereas a recent study \cite{huapeskin} analyzed stability for a spatially discretized boundary \cite{huapeskin} and a particular regularized delta function, the analog of the blob function which communicates between the Eulerian/Lagrangian grids in IB methods. We note that since we will focus on the MRS, our stability analysis will be similar in flavor, but the linearization will be related to the regularized fundamental solution. 

 We first complete a linear stability analysis of a doubly periodic elastic surface immersed in a viscous fluid, modeled using the MRS. This allows us to formulate an eigenvalue problem from which a time step restriction can be directly determined. For the case of an algebraic blob function (given in \eqref{eq:algblob}) and a tensile model of the elastic surface, we illustrate how the decay rates of the higher wavenumbers grow as the regularization parameter $\epsilon$ decreases relative to the discretization spacing $h$ for the elastic structure. We present theoretical results for different blob functions. These results use newly derived doubly periodic regularized Stokeslets for three blob functions which are recorded in the appendices. We numerically verify the results with simulations based on the nonlinear system and highlight how there is a tradeoff between minimizing regularization error in the far field (away from the structure) and the time step restriction. Similarly, we show that when accounting for bending and tension, the time step restriction is dominated by the bending model. We illustrate how this analysis in a doubly periodic domain can provide guidance in the choice of regularization function, regularization parameter, and time step for commonly used elastic forcing models for the case of a finite elastic surface in free space. This example illustrates how theoretical relationships derived in the doubly periodic setting can be utilized as a guide for the practitioner even if simulating elastic structures in different domains.

\section{Coupled Elastic Surface-Fluid System}

\subsection{Elastic Surface}
The elastic surface is immersed in the fluid and denoted as   $\vec X(q,s,t)$ where $t$ is time and Lagrangian coordinates $(q,s)$ correspond to an equilibrium reference configuration of the structure lying flat in the $z=0$ plane ($\vec X^{eq}(q,s) = (q,s,0)$) with $0\leq q,s \leq L$. 
When the surface is perturbed from equilibrium, it exerts forces on the fluid due to its tensile rigidity and bending rigidity. 
The tension of the material in the first (second) material direction is denoted $T^q$ ($T^s$). We model the tension in one direction as a linear function of the strain in that direction, i.e. 
\begin{equation}
    \label{eq:tension}
    T^q = T^q \left ( \left | \pp{\vec X}{q} \right | \right ) = \sigma_q  \left | \pp{\vec X}{q} \right |,   \ \ \ T^s = T^s \left( \left | \pp{\vec X}{s} \right | \right ) = \sigma_s \left | \pp{\vec X}{s} \right |,
\end{equation}
where $\sigma_q, \sigma_s$ are constants with dimensions force per unit length.
Using variational principles \cite{fauci1988, peskin}, it follows that the surface force density from tension at an arbitrary point on the structure is \begin{equation}
    \vec F_T (\vec X(q,s,t) ) = \sigma_q \pptwo{\vec X}{q} + \sigma_s \pptwo{\vec X}{s}.
    \label{eq:tensionForce}
\end{equation}
In a similar fashion, we can define a linear variant of the bending-resistant surface force density \cite{Landau1970} 
as
\begin{equation}
    \vec F_B (\vec X(q,s,t) ) = -\kappa_b \left ( \frac{\partial^4 \vec X}{\partial q^4} + 2\frac{\partial^4 \vec X}{\partial q^2 \partial s^2} + \frac{\partial^4 \vec X}{\partial s^4} \right).
\label{eq:bendingForce}
\end{equation}where $\kappa_B$ has dimensions force times length. 
 The tension and bending-resistant forces are both zero when the structure is flat, i.e. it is in the reference configuration $\vec X^{eq}$. When only accounting for tension, $\vec F$ 
is set to $\vec F=\vec F_T$ and when accounting for both tension and bending, we set $\vec F=\vec F_T+\vec F_B$ (where $\vec F_T$ is given in \eqref{eq:tensionForce} and $\vec F_B$ is given in \eqref{eq:bendingForce}).

 \subsection{Regularized Boundary Integral Equations}\label{sec:MRS}

Solutions to the Stokes equations in 3D involving immersed bodies or surfaces can be represented as singular boundary integral equations relating the fluid velocity $\vec u$ and the surface force density $\vec F$ \cite{pozr1992}. Analogously, one can derive a regularized version of the boundary integral equations. For simplicity, consider a single structure $\Gamma_t$ immersed in an unbounded Stokes fluid. Then the exact velocity at a point $\vec x$ due to the presence of the structure, $\vec u(\vec x)$, can be approximated as \cite{cortez2005, smithDouble},
\begin{align}
 u_j(\vec x)&\approx \iiint_{\mathbb{R}^3}u_j(\vec x, \vec y)\phi_{\epsilon}(\vec x, \vec y) dV(\vec y)  \nonumber \\
 &=\frac{1}{8 \pi \mu}\iint_{\Gamma_t} \mathcal{S}_{\phi_\epsilon,ij}(\vec x, \vec X) F_i(\vec X) dS(\vec X)  \nonumber \\
 & +\frac{1}{8\pi}\iint_{\Gamma_t}u_i(\vec x)\mathcal{T}_{\phi_\epsilon, ijk}(\vec x,\vec X)n_k(\vec X) dS(\vec X), \label{eq:regboundint}
 \end{align}
where $\mathcal{S}_{\phi_\epsilon,ij}$ is the regularized Stokeslet, $\mathcal{T}_{\phi_\epsilon,ijk}$ is the regularized stress tensor, $\vec n$ is the unit outward normal vector to $\Gamma_t$,  and repeated indices utilize the standard summation convention. The first double integral in \eqref{eq:regboundint} is called the single-layer potential, and the second double integral is called the double-layer potential. It is sometimes reasonable to eliminate the double-layer potential in \eqref{eq:regboundint}. One such case is a thin sheet which is idealized to have zero thickness. To see this, consider a thin sheet with thickness $\varepsilon_m$. In the limit as $\varepsilon_m \to 0$, the continuity of the velocity field across the boundary and the opposite sign of the normal vector $\vec n$ cancels out the contribution from either side \cite{pozr2011}. Hence, we will neglect the double layer potential in \eqref{eq:regboundint}. We note that such a simplification is not always possible and there are cases where the double layer potential may be of importance. We refer the reader to a commentary by Smith \cite{smithDouble} which discusses how to handle this integral within the framework of the MRS. 

The boundary integral formulation and the simplification to the single-layer potential allow us to write the solution of our coupled elastic surface-fluid system with regularized forces formally as

\begin{equation}
    \begin{split}
    \vec u (\vec x) &= \frac{1}{8\pi\mu}\iint_{\Gamma_t} \Stokes (\vec x, \vec X(q,s,t) ) \cdot \vec F(\vec X(q,s,t)) dS(q,s), \\ 
    \frac{\partial{\vec X(q,s,t)}}{\partial t} & = \vec u (\vec X(q,s,t)).
    \end{split}
    \label{eq:nonlinearSystem}
\end{equation}Note that this system of equations nonlinearly couples the fluid velocity $\vec u$ and the surface $\vec X$. The first equation says that at every instant $t$, the fluid velocity is determined by the surface configuration at that time. The second equation is the no-slip condition which matches the structure velocity with the fluid velocity and effectively evolves the system over time. 

Details of the numerical implementation of the MRS are given in \ref{app:regstokes} (free space) and \ref{app:ffts} (doubly periodic). We will focus on the regularized boundary integral formulation in \eqref{eq:nonlinearSystem}, for a given regularized Stokeslet $\boldsymbol{\mathcal{S}}_{\phi_\epsilon}$ that satisfies the appropriate boundary conditions, whether in free-space or a doubly periodic domain.

\section{Fourier Analysis of Linearized System} 
\label{sec:fourier}
\begin{figure}[htb!]
        \centering   
        \includegraphics[width=0.75\textwidth]{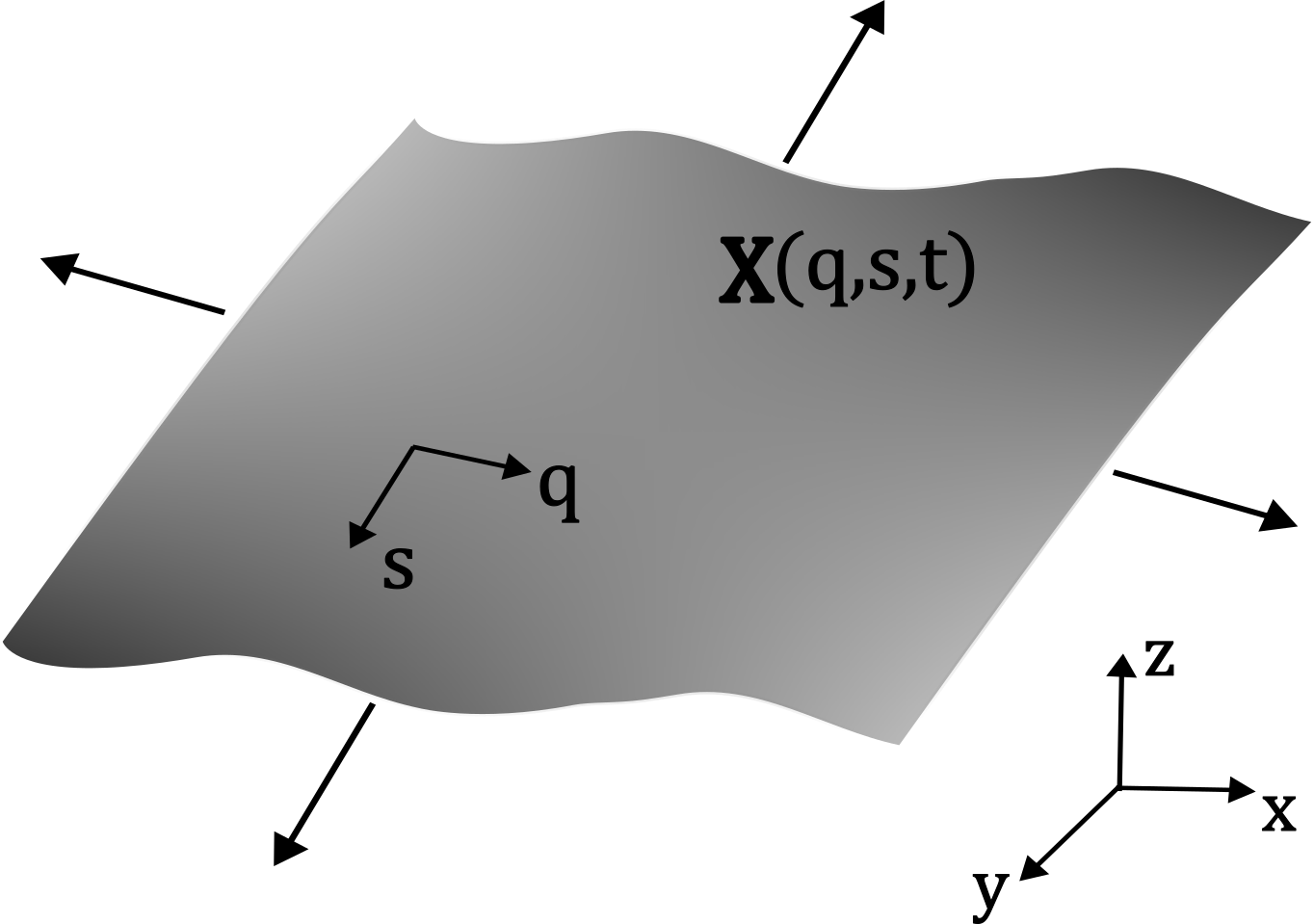}
        \caption{Diagram of the doubly periodic elastic surface $\vec X(q,s,t)$ with $q,s$ material coordinates in the flat reference configuration, $\vec X^{eq}(q,s)=(q,s,0)$. Arrows indicate the periodic extension of the surface in the $q,s$ directions. }
        \label{fig:surface}
\end{figure}

To analyze the stability of the coupled elastic surface-fluid system in \eqref{eq:nonlinearSystem}, we consider the problem in a doubly periodic domain. We let $x,y$ be the periodic directions with length $L$ and $z$ the free direction. This constrains the domain of interest to $(x,y,z) \in [0,L] \times [0,L] \times \mathbb{R}.$ For simplicity, we assume the Lagrangian coordinates $q,s$ are parameters of arc-length in the equilibrium configuration so that the solution for $\vec X(q,s,t)$ is also periodic with length $L$ (see Fig.~\ref{fig:surface}). The periodic assumptions will make the tools of Fourier analysis applicable. 

\subsection{Separable Solutions - Continuous Formulation}\label{sec:ansatz}
As is standard in linear stability analyses, we assume a small perturbation is applied to the surface's equilibrium configuration, $\vec X^{eq}(q,s)$, and use the \textit{ansatz} for the velocity $\vec u$ and structure position $\vec X$,

\begin{align}
    \vec u(\vec x; t) &= \sum_{\alpha, \beta\in \mathbb{Z}} \exp \left ( \frac{2 \pi i}{L} \left ( \alpha x  + \beta y \right )  \right ) \whvec{u}_{\alpha,\beta}(z;t), \label{eq:ufourier} \\
    \vec X(q, s, t)  &= \sum_{\alpha, \beta\in \mathbb{Z}} \exp \left ({\lambda_{\alpha,\beta} t + \frac{2 \pi i}{L}\left ( \alpha q  + \beta s \right ) } \right ) \whvec{X}_{\alpha, \beta}, \label{eq:Xfourier}
\end{align}
where $i=\sqrt{-1}$ and $\mathbb{Z}$ is the set of integers. We interpret $\lambda_{\alpha,\beta}=\lambda({\alpha, \beta})$ as the growth or decay rate of the $(\alpha, \beta)$ Fourier mode. Note that $\vec u$ depends on time only through the surface position $\vec X(q,s,t)$, as indicated by the way the arguments of $\vec u $ are written. The Fourier coefficients $\whvec{u}_{\alpha, \beta}(z)$ and $\whvec X_{\alpha, \beta}$ are respectively 

\begin{align}
    \whvec{u}_{\alpha, \beta}(z; t) &= \frac{1}{L^2} \iint_{R_L} \vec u(\vec x;t) \exp{\left(\frac{-2 \pi i}{L} \left(\alpha x + \beta y \right) \right ) \ dx \ dy}, 
    \label{eq:ualpha} \\
    \whvec{X}_{\alpha, \beta}  &= \frac{1}{L^2} \iint_{R_L} \vec X(q,s,t) \exp{\left(\frac{-2 \pi i}{L} \left(\alpha q + \beta s \right) \right ) \ dq \ ds}, \label{eq:Xalpha}
\end{align}
where $R_L$ is the rectangular domain $[0,L] \times [0,L]$. Since the forces on the structure are a function of the surface position given by $\vec X$, we can rewrite the tension force in \eqref{eq:tensionForce} as a Fourier series, 

 \begin{equation}
     \begin{split}
     \label{eq:ffourier}
      \vec F_T(\vec X(q,s,t) ) &= \sigma_q \pptwo{\vec X}{q} + \sigma_s \pptwo{\vec X}{s} \\
      &=  \frac{-4 \pi^2}{L^2} \sum_{\alpha, \beta\in \mathbb{Z}} \left (\sigma_q \alpha^2 + \sigma_s \beta^2 \right ) \exp \left ({\lambda_{\alpha,\beta} t + \frac{2 \pi i}{L}\left ( \alpha q  + \beta s \right ) } \right ) \whvec{X}_{\alpha, \beta}.   
      \end{split}
 \end{equation}Similarly, the bending-resistant force is 

 \begin{equation}
     \begin{split}
     \label{eq:fbfourier}
      \vec F_B(\vec X(q,s,t) ) &= -\kappa_b \left ( \frac{\partial^4 \vec X}{\partial q^4} + 2\frac{\partial^4 \vec X}{\partial q^2 \partial s^2} + \frac{\partial^4 \vec X}{\partial s^4} \right) \\
      &= - \frac{16 \pi^4}{L^4}  \sum_{\alpha, \beta\in \mathbb{Z}} \kappa_B \left (\alpha^2 + \beta^2 \right )^2 \exp \left ({\lambda_{\alpha,\beta} t + \frac{2 \pi i}{L}\left ( \alpha q  + \beta s \right ) } \right ) \whvec{X}_{\alpha, \beta}.  
      \end{split}
 \end{equation} The general case of a surface with both tensile and bending resistant forces is compactly written as

  \begin{equation}
     \begin{split}
     \label{eq:ftbfourier}
      \vec F(\vec X(q,s,t)) &= \left [ \vec F_T + \vec  F_B \right ] (\vec X(q,s,t) ) \\ 
      &= \frac{1}{L^2}\sum_{\alpha, \beta \in \mathbb{{Z}}}K_{\alpha, \beta} \exp \left ({\lambda_{\alpha,\beta} t + \frac{2 \pi i}{L}\left ( \alpha q  + \beta s \right ) } \right ) \whvec{X}_{\alpha, \beta}, \\ 
K_{\alpha, \beta} &=  - 4 \pi^2\left (\sigma_q \alpha^2 + \sigma_s \beta^2 \right ) - \frac{16 \pi^4}{L^2}\kappa_B \left (\alpha^2 + \beta^2 \right )^2.  
      \end{split}
 \end{equation}

 We utilize the regularized boundary integral formulation in \eqref{eq:nonlinearSystem} to determine the fluid velocity due to the surface forces on the structure. Since we are in a doubly periodic domain, we have 
 \begin{equation}
\label{eq:singleLayer}
\vec u (\vec x; t) = \frac{1}{8 \pi \mu} \iint_{\Gamma_t} \boldsymbol{\mathcal{S}}^{DP}_{\phi_{\epsilon}}(\vec x, \vec X(q,s,t)  )
\cdot \vec F(\vec X(q,s,t) ) \ dS(q,s), 
\end{equation}where $\boldsymbol{\mathcal{S}}^{DP}_{\phi_\epsilon}$ is the doubly periodic (DP) version of the regularized Stokeslet, 

\begin{equation}
    \label{eq:SDP}
\boldsymbol{\mathcal{S}}^{DP}_{\phi_{\epsilon}}(\vec x, \vec y ) = \sum_{\alpha, \beta\in \mathbb{Z}} \widehat{\boldsymbol{\mathcal{S}}}^{DP}_{\phi_{\epsilon}}
    (r_3; \alpha, \beta) \exp{ \left( \frac{2 \pi i}{L} \left (r_1 \alpha + r_2 \beta\right )\right )},
\end{equation}
for $\vec r= \vec x - \vec y$. The derivation of the Fourier coefficients $\widehat{\boldsymbol{\mathcal{S}}}^{DP}_{\phi_{\epsilon}}(r_3; \alpha, \beta)$ will be detailed in Section \ref{sec:blobChoice}. 

\subsection{Linear Approximation to Obtain the Eigenvalue Problem}
The integrand in \eqref{eq:singleLayer} is a nonlinear function of $\vec X(q,s,t)$. Since the perturbation to the surface is assumed to be small, $|\vec X(q,s,t)- \vec X^{eq}(q,s))| \ll 1$, we approximate the doubly periodic regularized Stokeslet in \eqref{eq:singleLayer} as
\begin{equation}
    \label{eq:stokesletLinear}
    \Sdp(\vec x, \vec X(q,s,t))  \approx \Sdp (\vec x, \vec X^{eq}(q,s)).
\end{equation}Substituting \eqref{eq:ftbfourier}, \eqref{eq:SDP}, \eqref{eq:stokesletLinear}, and the differential area approximation $dS(q,s) \approx dq \ ds$ into \eqref{eq:singleLayer} yields

\begin{equation}
 \begin{split}        
     \vec u(\vec x; t) &= \frac{1}{8 \pi \mu} \iint_{R_L} \left (\sum_{\alpha', \beta'\in\mathbb{Z}} \Sdp (z; \alpha', \beta') \exp \left (\frac{2 \pi i}{L} \left ( (x-q) \alpha' + (y-s) \beta' \right ) \right ) \right )\cdot  \\
     &\frac{1}{L^2}\left[ \sum_{\tilalpha, \tilbeta\in\mathbb{Z}} K_{\tilalpha,\tilbeta} \exp \left ({\lambda_{\tilalpha, \tilbeta} t + \frac{2 \pi i}{L}\left ( \tilalpha q  + \tilbeta s \right ) } \right ) \whvec{X}_{\tilalpha, \tilbeta} \right ] dq \ ds. \  \end{split}\label{eq:singlelayer1}
 \end{equation}
 This equation is a linearized approximation of the boundary integral equation \eqref{eq:singleLayer}. Next, we multiply each side by $\exp{\left ( -\frac{2 \pi i}{L} (\alpha x + \beta y)  \right )}$ where $\alpha, \beta$ are \textit{fixed} integers and integrate each side with respect to $x, y$ over $R_L$. After rearrangement of terms and changing the order of integration on the right hand side, we have

 \begin{equation} 
 \begin{split}
     \label{eq:uStepOne}
     \whvec{u}_{\alpha, \beta}(z) = \frac{1}{8 \pi \mu} &\iint_{R_L}  \Sdp (z; \alpha', \beta') \exp \left (\frac{-2 \pi i}{L}(q \alpha + s \beta) \right ) \cdot \\ 
     &\frac{1}{L^2}\left[ \sum_{\tilalpha, \tilbeta} K_{\tilalpha, \tilbeta}
      \exp \left ({\lambda_{\tilalpha, \tilbeta} t + \frac{2 \pi i}{L}\left ( \tilalpha q  + \tilbeta s \right ) } \right ) \whvec{X}_{\tilalpha, \tilbeta} \right ] dq \ ds .
     \end{split}
 \end{equation}The right hand side is obtained by using the identity 

 \begin{equation}
     \label{eq:expOrthog}
     \iint_{R_L} \exp \left ( \frac{2 \pi i}{L} \left ( x(\alpha'-\alpha) + y(\beta' - \beta) \right)  \right ) \ dx \ dy = \begin{cases}
     L^2, & \alpha' =\alpha, \ \beta' = \beta \\
     0, & \text{otherwise}
 \end{cases}.
 \end{equation}The left hand side of \eqref{eq:uStepOne} is obtained by using \eqref{eq:ualpha}. Removing the common factor $L^2$ from each side results in the final form of \eqref{eq:uStepOne}. To further simplify, we rewrite \eqref{eq:uStepOne} as
 \begin{equation} 
 \begin{split}
     \label{eq:uStepTwo}
     \whvec{u}_{\alpha, \beta}(z) &= \frac{1}{8 \pi \mu} \iint_{R_L}  \Sdp  (z; \alpha, \beta) \cdot \\ 
     &\frac{1}{L^2}\left[  \sum_{\tilalpha, \tilbeta} K_{\tilalpha, \tilbeta}
      \exp \left ({\lambda_{\tilalpha, \tilbeta} t + \frac{2 \pi i}{L}\left ( (\tilalpha - \alpha) q  + (\tilbeta - \beta) s \right ) } \right ) \whvec{X}_{\tilalpha, \tilbeta} \right ] dq \ ds .
     \end{split}
 \end{equation}Using \eqref{eq:expOrthog}, \eqref{eq:uStepTwo} simplifies to
  \begin{equation} 
     \label{eq:uStepThree}
     \whvec{u}_{\alpha, \beta}(z) = \frac{1}{8 \pi \mu}  \Sdp (z; \alpha, \beta) \cdot \left [ K_{\alpha, \beta}\exp \left ({\lambda_{\alpha, \beta} t } \right )\whvec{X}_{\alpha, \beta}  \right ].
 \end{equation}
 
 The no-slip condition, $\pp{\vec X(q,s,t)}{t}=\vec u(\vec X(q,s,t))$, is another source of nonlinearity in our system. The linearized approximation is

 \begin{equation}
    \label{eq:linapprox2} 
  \pp{\vec X(q,s,t)}{t} = \vec u \left( \vec X(q,s,t) \right) \approx \vec u \left( \vec X^{eq}(q,s); t \right) = \vec u (q,s,0).
\end{equation}
Using the Fourier representation of $\vec X$ in \eqref{eq:Xfourier}, the surface velocity is 

 \begin{equation} 
     \label{eq:dXdt}
     \pp{\vec X(q,s,t)}{t} = \sum_{\alpha, \beta} \lambda_{\alpha, \beta} \exp \left ({\lambda_{\alpha, \beta} t + \frac{2 \pi i}{L}\left ( \alpha q  + \beta s \right ) } \right ) \whvec{X}_{\alpha, \beta}.
 \end{equation}With the Fourier representation of $\vec u$ in \eqref{eq:ufourier} and the Fourier representation of $\pp{\vec X(q,s,t)}{t}$ in \eqref{eq:dXdt}, we substitute into \eqref{eq:linapprox2},

 \begin{equation} 
 \begin{split}
     \label{eq:interfaceFourier}
     \sum_{\alpha', \beta'} &\exp \left ({\lambda_{\alpha',\beta'} t + \frac{2 \pi i}{L} \left ( \alpha' q  + \beta' s \right ) } \right ) \whvec{u}_{\alpha',\beta'}(0;t)  \\ 
     &= \sum_{\tilalpha, \tilbeta} \lambda_{\tilalpha, \tilbeta} \exp \left ({\lambda_{\tilalpha, \tilbeta} t + \frac{2 \pi i}{L}\left ( \tilalpha q  + \tilbeta s \right ) } \right ) \whvec{X}_{\tilalpha, \tilbeta}.
     \end{split}
 \end{equation}By the discrete orthogonality of the complex exponentials, we can simplify \eqref{eq:interfaceFourier} as

 \begin{equation}
     \label{eq:uStepFour}
     \whvec {u}_{\alpha, \beta} (0) = \lambda_{\alpha, \beta}\whvec{X}_{\alpha, \beta} \exp { \left ( \lambda_{\alpha, \beta} t \right )}
 \end{equation}for every $\alpha, \beta$. Finally, we equate the right hand side of \eqref{eq:uStepFour} and the right hand side of \eqref{eq:uStepThree} evaluated at $z=0$. After canceling the common exponential factors, we have

 \begin{equation}
     \label{eq:eigenvalueProblem}
      \lambda_{\alpha, \beta} \whvec{X}_{\alpha, \beta} =  \frac{K_{\alpha, \beta}}{8 \pi \mu}  \Sdp  (0; \alpha, \beta) \cdot \whvec{X}_{\alpha, \beta}.
 \end{equation} 
Equation \eqref{eq:eigenvalueProblem} is an eigenvalue problem and a particular solution requires a choice of blob function, $\phi_\epsilon$. 
The derivation for the general form of the doubly periodic regularized Stokeslet $\boldsymbol{\mathcal{S}}^{DP}_{\phi_\epsilon}$ is in Section \ref{sec:blobChoice}, and specific formulas for particular blob functions are reported in \ref{app:DPSB} and \ref{app:blobmoment}.

 \subsection{Eigenvalue Problem for Spatially Discretized System}\label{sec:evalue}
In practice, the forces that the structure exerts on the fluid involve the approximation of spatial derivatives on the immersed boundary and a quadrature rule to approximate the boundary integral for the velocity in \eqref{eq:singleLayer}. However, a discretization of the fluid domain is not necessary in most implementations of the MRS since the fluid velocity is determined by the forces along the boundary at any instant in time (we have a solution that can be evaluated on or off the structure). Hence, we will introduce a discretization of the surface and the boundary integral, but retain the continuous representation of the fluid velocity. 

To discretize the surface, we use $N$ grid points in each material direction: $(q_j, \ s_k) = (jh, \ kh)$ with $h=L/N$. For notational simplicity, we take $N$ to be even. Since the surface is doubly periodic, we restrict $j,k \in \Zmodn=\{0,1, \dots, N-1\}$. Conversely, we restrict the set of discrete Fourier wavenumbers $\alpha, \beta \in \Fmodn = \{-N/2, -N/2+1, \dots, 0, \dots, N/2-1\}$. The discretized surface position $\vec X(q_j, s_k, t)$ is now represented as

\begin{align}
    \vec X(q_j, s_k, t)  &= \sum_{\alpha, \beta\in \Fmodn} \exp \left ({\lambda_{\alpha, \beta} t + \frac{2 \pi i}{L}\left ( \alpha q_j  + \beta s_k \right ) } \right ) \whvec{X}_{\alpha, \beta}, \label{eq:XfourierDiscrete} \\ 
     \whvec{X}_{\alpha, \beta}  &= \frac{1}{L^2} \sum_{j,k \in \Zmodn} \exp{\left(\frac{-2 \pi i}{L} \left(\alpha q_j + \beta s_k \right) \right ) \vec X(q_j,s_k,t) h^2}. \label{eq:XalphaDiscrete}
\end{align}The representation of the fluid velocity $\vec u$ and the Fourier coefficients $\whvec u_{\alpha,\beta}$ remain the same as \eqref{eq:ufourier} and \eqref{eq:ualpha} respectively, with the only difference being that the wavenumbers $\alpha,\beta$ are now restricted to $\Fmodn$.  

To determine the discretized version of the tension force, we have to rewrite the Fourier series expansion in \eqref{eq:ffourier}. Recall that the discrete, centered second derivative operator for a function $f=f(x_j)$ defined on a grid $x_j=jh$ with uniform spacing $h=L/N$ is 
$D^2 f(x_j) =\frac{N}{L^2}(f(x_{j+1})- 2 f(x_j)+f(x_{j-1}))$.
Using the Fourier representation $f(x_j) = 
\sum_{\alpha\in\Fmodn} \exp{\left(\frac{2 \pi i}{L} \alpha x_j\right ) \widehat{f}(\alpha)}$, it can be shown that 
\begin{equation}
\begin{split}
    \label{eq:D2fourier}
    D^2 f(x_j) &= \sum_{\alpha\in\Fmodn} \widehat{D}^2(\alpha)\exp{\left(\frac{2 \pi i}{L} \alpha x_j\right ) \widehat{f}(\alpha)}  \\
    &= \frac{-4N^2}{L^2}\sum_{\alpha\in\Fmodn} \left (\sin^2 \frac{\pi \alpha h}{L}  \right )\exp{\left(\frac{2 \pi i}{L} \alpha x_j\right ) \widehat{f}(\alpha)}. 
    \end{split}
\end{equation}Applying \eqref{eq:D2fourier} component-wise on the surface $\vec X$, the continuous tension force \eqref{eq:ffourier} can now be discretized as 

 \begin{equation}
     \begin{split}
     \label{eq:fdiscrete}
\vec F_{T_{j,k}}&= \sigma_q D_q^2 \vec X(q_j,s_k,t) + \sigma_s D_s^2 \vec X(q_j,s_k,t) \\
      &= \frac{-4N^2}{L^2}  \sum_{\alpha, \beta\in \Fmodn} \left ( \sigma_q \sin^2 \frac{\pi \alpha h}{L} + \sigma_s \sin^2 \frac{\pi \beta h}{L} \right ) \cdot \\ 
      &\hspace{6em} \exp \left ({\lambda_{\alpha, \beta} t + \frac{2 \pi i}{L}\left ( \alpha q_j  + \beta s_k \right ) } \right ) \whvec{X}_{\alpha, \beta},   
      \end{split}
 \end{equation}where the subscripts on the finite difference operators indicate the material direction that is being operated on. The discrete analog of the continuous bending force \eqref{eq:ftbfourier} likewise becomes

 \begin{equation}
     \begin{split}
     \label{eq:fbdiscrete}
\vec F_{B_{j,k}}&= - \kappa_B (D^2_q + D^2_s)(D^2_q + D^2_s) \vec X(q_j, s_k, t) \\
      &= \frac{-16N^4}{L^4} \kappa_B  \sum_{\alpha, \beta\in \Fmodn} \left (\sin^2 \frac{\pi \alpha h}{L} + \sin^2 \frac{\pi \beta h}{L} \right )^2 \cdot \\ 
       & \hspace{8em} \exp \left ({\lambda_{\alpha, \beta} t + \frac{2 \pi i}{L}\left ( \alpha q_j  + \beta s_k \right ) } \right ) \whvec{X}_{\alpha, \beta}.   
      \end{split}
 \end{equation}As before, we can rewrite the general force term as 

 \begin{equation}
     \label{eq:ftbdiscrete}
     \begin{split}
     \vec F_{j,k} &= \vec F_{T_{j,k}} + \vec F_{B_{j,k}} \\
     &=\frac{1}{L^2} \sum_{\alpha, \beta \in \Fmodn} \Kd \exp \left ({\lambda_{\alpha, \beta} t + \frac{2 \pi i}{L}\left ( \alpha q_j  + \beta s_k \right ) } \right ) \whvec{X}_{\alpha, \beta},  \\ 
\Kd &= -4N^2 \left ( \sigma_q \sin^2 \frac{\pi \alpha h}{L} + \sigma_s \sin^2 \frac{\pi \beta h}{L} \right ) \\
     &\hspace{6em} - \frac{16N^4}{L^2}\kappa_B \left (\sin^2 \frac{\pi \alpha h}{L} + \sin^2 \frac{\pi \beta h}{L} \right )^2. 
     \end{split}
 \end{equation}
 
Using the same linear approximation in \eqref{eq:stokesletLinear}, we discretize the boundary integral over the surface structure from \eqref{eq:singlelayer1} to determine the velocity, $\vec u(\vec x;t)$. Utilizing the discrete force term \eqref{eq:ftbdiscrete} and replacing the integral $\iint_{R_L} \ dq \ ds$ with a summation over the grid with weighting $h^2$, we have

 \begin{equation}
 \begin{split}
     \label{eq_j:singlelayerdiscrete}    
     \vec u(\vec x; t) = \frac{1}{8 \pi \mu} &\sum_{j,k \in \Zmodn} \Bigg [ \sum_{\alpha', \beta'\in \Fmodn} \Sdp (z; \alpha', \beta') \cdot \\ 
     &\exp \left (\frac{-2 \pi i}{L}(q_j \alpha' + s_k \beta') \right ) 
     \exp \left (\frac{2 \pi i}{L}(x \alpha' + y \beta') \right ) \cdot \\ 
     &\frac{1}{L^2}  \sum_{\tilalpha, \tilbeta\in \Fmodn} \Kd \exp \left ({\lambda_{\tilalpha, \tilbeta} t + \frac{2 \pi i}{L}\left ( \tilalpha q_j  + \tilbeta s_k \right ) } \right ) \whvec{X}_{\tilalpha, \tilbeta} \ h^2 \Bigg ].  
      \end{split}
 \end{equation}
 
 The goal as before is to obtain an eigenvalue problem.  As we did in the continuous problem, we multiply each side of \eqref{eq_j:singlelayerdiscrete} by $\exp \left ( -\frac{2 \pi i} {L} \left ( \alpha x + \beta y\right )\right )$ where $\alpha, \beta \in \Fmodn$ are fixed integers. We then integrate each side with respect to $x,y$ over $R_L$. By switching the order of the summation with respect to $\alpha', \beta'$ with the integration over $R_L$ and applying the identity \eqref{eq:expOrthog}, we have
 \begin{equation} 
 \begin{split}
     \label{eq:fourierStepOne}
     \whvec{u}_{\alpha, \beta}(z) &= \frac{1}{8 \pi \mu} \sum_{j, k \in \Zmodn}  \whvec {\boldsymbol{\mathcal{S}}}^{DP}_{\phi_{\epsilon}}  (z; \alpha, \beta) \cdot \\ 
     &\left[ \frac{1 }{L^2}  \sum_{\tilalpha, \tilbeta\in \Fmodn } \Kd 
      \exp \left ({\lambda_{\tilalpha, \tilbeta} t + \frac{2 \pi i}{L}\left ( (\tilalpha - \alpha) q_j  + (\tilbeta - \beta) s_k \right ) } \right ) \whvec{X}_{\tilalpha, \tilbeta} \right ] \ h^2.
     \end{split}
 \end{equation}Now we switch the order of the summations and use the discrete form of the identity \eqref{eq:expOrthog}, 
 
 \begin{equation}
     \label{eq:discExpOrthog} 
     \sum_{j, k \in \Zmodn} \exp \left ( \frac{2 \pi i}{L} \left ( q_j(\tilalpha-\alpha) + s_k(\tilbeta - \beta) \right)  \right ) \ h^2 = \begin{cases}
     L^2, & \tilalpha =\alpha, \ \tilbeta = \beta \\
     0, & \text{otherwise}
     \end{cases}
 \end{equation}so that the right hand side collapses into 

 \begin{equation} 
     \begin{split}
     \whvec{u}_{\alpha, \beta}(z) = &\frac{1}{8 \pi \mu} \whvec {\boldsymbol{\mathcal{S}}}^{DP}_{\phi_{\epsilon}}  (z; \alpha, \beta) \cdot \left [\Kd  \exp \left (\lambda_{\alpha,\beta} t \right )\whvec X_{\alpha, \beta} \right ].
     \end{split}
 \end{equation}
 
 Applying the linear approximation \eqref{eq:linapprox2} and following the same arguments detailed in the previous section using \eqref{eq:dXdt}, \eqref{eq:interfaceFourier}, and \eqref{eq:uStepFour}, we obtain an eigenvalue problem for the spatially discretized system, 
\begin{equation}
    \label{eq:eigenvalueProblem2}
     \lambda_{\alpha, \beta} \whvec{X}_{\alpha, \beta} =  \frac{\Kd}{8 \pi \mu} \whvec {\boldsymbol{\mathcal{S}}}^{DP}_{\phi_{\epsilon}}  (0; \alpha, \beta) \cdot \whvec X_{\alpha, \beta} .
 \end{equation} \subsection{Doubly Periodic Regularized Stokeslet $\boldsymbol{\mathcal{S}}^{DP}_{\phi_{\epsilon}}$} \label{sec:blobChoice}
The eigenvalue problem for the  linearized elastic surface-fluid system depends on the doubly periodic regularized Stokeslet $\boldsymbol{\mathcal{S}}^{DP}_{\phi_{\epsilon}}$ in both the continuous and discrete cases, \eqref{eq:eigenvalueProblem} and \eqref{eq:eigenvalueProblem2}, respectively. To determine $\boldsymbol{\mathcal{S}}^{DP}_{\phi_{\epsilon}}$, we first need to specify the blob $\phi_{\epsilon}$. Following the method of \cite{hoffmann1}, the first piece that is required is the doubly periodic extension of $\phi_{\epsilon}$: 

\begin{align}     
    \phi^{DP}_{\epsilon}(\vec r) &= \sum_{\alpha, \beta\in\mathbb{Z}_N} \widehat{\phi}^{DP}_{\epsilon}(z; \alpha, \beta) \exp {\left ( \frac{2 \pi i}{L} \left ( \alpha r_1 + \beta r_2\right) \right )}, \label{eq:phi2D}\\
    \widehat{\phi}_{\epsilon}^{DP}(z; \alpha, \beta) &= \frac{1}{L^2} \iint_{\R^2} \phi_\epsilon(\vec r)\exp {\left ( - \frac{2 \pi i}{L} \left ( \alpha r_1 + \beta r_2 \right) \right )} \ dr_1 \ dr_2 . \label{eq:phiFT}
\end{align}Here, \eqref{eq:phiFT} is the 2D Fourier transform of $\phi_\epsilon$ scaled by $L^{-2}$. Assuming \eqref{eq:phi2D}-\eqref{eq:phiFT} are well defined, we seek functions $\widehat{G}^{DP}_{{\phi}_{\epsilon}}(z; \alpha, \beta), \widehat{B}^{DP}_{{\phi}_{\epsilon}}(z; \alpha, \beta)$ that solve the following ordinary differential equations in $z$,
\begin{align}
    \widehat{\Delta} \widehat{G}^{DP}_{{\phi}_{\epsilon}}(z; \alpha, \beta) &= \widehat{\phi}^{DP}_{\epsilon}(z; \alpha, \beta) \label{eq:Gode} \\
    \widehat{\Delta} \widehat{B}^{DP}_{{\phi}_{\epsilon}}(z; \alpha ,\beta) &= \widehat{G}^{DP}_{{\phi}_{\epsilon}}(z; \alpha, \beta) \label{eq:Bode} \end{align}
where $\widehat{\nabla} = \left( \frac{2 \pi i}{L} \alpha, \frac{2 \pi i}{L} \beta, \frac{d}{dz} \right ), \ \widehat{\Delta} = \widehat{\nabla} \cdot \widehat{\nabla}$. 

To ensure that $|\vec u| \to 0$ as $z \to \pm \infty$, one may need to utilize solutions to the associated homogeneous ODEs in $z$ for $\widehat{G}^{DP}$ and $\widehat{B}^{DP}$ as in \cite{hoffmann1}. Once $\widehat{G}^{DP}_{\phi_{\epsilon}}, \widehat{B}^{DP}_{\phi_{\epsilon}}$ are known, the Fourier coefficients $\whvec{\boldsymbol{\mathcal{S}}}^{DP}_{\phi_{\epsilon}}(z; \alpha, \beta)$ can be found by evaluating

\begin{equation}
\label{eq:StokesletForm}
     \whvec{\boldsymbol{\mathcal{S}}}^{DP}_{\phi_{\epsilon}}(z; \alpha, \beta) = 8 \pi \left [ \widehat{\nabla} \widehat{\nabla}  \widehat{B}^{DP}_{{\phi}_{\epsilon}}(z; \alpha, \beta) - \mathbb{I} \widehat{G}^{DP}_{{\phi}_{\epsilon}}(z; \alpha, \beta)\right ] \end{equation}
where $\mathbb{I}$ is the $3 \times 3$ identity matrix. The components of \eqref{eq:StokesletForm} are

\begin{equation} 
8 \pi 
   \begin{bmatrix}
        -\frac{4 \pi^2 \alpha^2}{L^2} \widehat{B}^{DP}_{{\phi}_{\epsilon}}-\widehat{G}^{DP}_{{\phi}_{\epsilon}} & - \frac{4 \pi^2 \alpha \beta}{L^2} \widehat{B}^{DP}_{{\phi}_{\epsilon}} & \frac{2 \pi i\alpha}{L} \frac{d \widehat{B}^{DP}_{{\phi}_{\epsilon}}}{dz} \\ 
        -\frac{4 \pi^2 \alpha \beta}{L^2} \widehat{B}^{DP}_{{\phi}_{\epsilon}} & - \frac{4 \pi^2 \beta^2}{L^2} \widehat{B}^{DP}_{{\phi}_{\epsilon}}-\widehat{G}^{DP}_{{\phi}_{\epsilon}} & \frac{2 \pi i\beta}{L} \frac{d\widehat{B}^{DP}_{{\phi}_{\epsilon}}}{dz} \\ 
        \frac{2 \pi i\alpha}{L} \frac{d \widehat{B}^{DP}_{{\phi}_{\epsilon}}}{dz} & \frac{2 \pi i \beta}{L} \frac{d\widehat{B}^{DP}_{{\phi}_{\epsilon}}}{dz} & \frac{4 \pi^2 }{L^2} (\alpha^2+\beta^2) \widehat{B}^{DP}_{{\phi}_{\epsilon}}
        \end{bmatrix} \ .
        \label{eq:Scoeff}
        \end{equation}

\subsection{Eigenmode Solutions for Continuous and Discrete Systems}
With the general form of the doubly periodic regularized Stokeslet, we can now solve for the eigenvalues and eigenvectors of the discrete eigenvalue problem derived at the end of Section \ref{sec:evalue}. Examining \eqref{eq:eigenvalueProblem} or \eqref{eq:eigenvalueProblem2}, it is apparent that the eigenvalues for either problem are the eigenvalues of $\widehat{\boldsymbol{\mathcal{S}}}^{DP}_{\phi_{\epsilon}}\left (0; \alpha, \beta \right )$, scaled by a multiplicative constant. The symmetry of \eqref{eq:Scoeff} also constrains the eigenvalues for every wavenumber $(\alpha, \beta)$ to be real.

It can be verified using the results of \ref{app:DPSB} and \ref{app:blobmoment} that $\frac{d \widehat B^{DP}_{\phi_{\epsilon}}}{dz} (0; \alpha, \beta)=0$ for $(\alpha, \beta) \neq (0,0)$ for all blobs considered. Simplifying $\widehat{\boldsymbol{\mathcal{S}}}^{DP}_{\phi_{\epsilon}}(0; \alpha, \beta)$, the Fourier coefficients of the doubly periodic regularized Stokeslet, given in \eqref{eq:Scoeff} and evaluated at $z=0$ for $(\alpha, \beta) \neq(0,0)$ are 
\begin{equation} 
8 \pi 
   \begin{bmatrix}
        -\frac{4 \pi^2 \alpha^2}{L^2} \widehat{B}^{DP}_{\phi_{\epsilon}}-\widehat{G}^{DP}_{\phi_{\epsilon}} & - \frac{4 \pi^2 \alpha \beta}{L^2} \widehat{B}^{DP}_{\phi_{\epsilon}} & 0 \\ 
        -\frac{4 \pi^2 \alpha \beta}{L^2} \widehat{B}^{DP}_{\phi_{\epsilon}} & - \frac{4 \pi^2 \beta^2}{L^2} \widehat{B}^{DP}_{\phi_{\epsilon}}-\widehat{G}^{DP}_{\phi_{\epsilon}} & 0\\ 
        0 & 0 & \frac{4 \pi^2 }{L^2} (\alpha^2+\beta^2) \widehat{B}^{DP}_{\phi_{\epsilon}}
        \end{bmatrix}
        \ .\label{eq:Scoef0}
        \end{equation}
\noindent The characteristic polynomial of \eqref{eq:Scoef0} 
       has three distinct roots, 
        \begin{equation}
            \tilde{\lambda}_1 = -\widehat{G}^{DP}_{\phi_{\epsilon}}, \hspace{.3cm} 
            \tilde{\lambda}_2 = -\widehat{G}^{DP}_{\phi_{\epsilon}} - \frac{4\pi^2}{L^2} (\alpha^2 + \beta^2) \widehat{B}^{DP}_{\phi_{\epsilon}}, \hspace{.3cm}  
            \tilde{\lambda}_3 = \frac{4\pi^2}{L^2} (\alpha^2 + \beta^2)  \widehat{B}^{DP}_{\phi_{\epsilon}} , \label{eq:lambdas}
        \end{equation}
and the associated eigenvectors are
\begin{equation}
       \begin{aligned}
        \tilde{\vec v}_1 &= \left[\frac{-\beta}{\alpha} \ \ 1 \ \ 0\right]^\top \ \ \mbox{for} \ \alpha \neq 0 \ \  \text{or } \ \tilde{\vec v}_1 = \left[ 1 \ \ \frac{-\alpha}{\beta} \ \ 0\right]^\top \ \ \mbox{for} \ \beta \neq 0,\\ \tilde{\vec v}_2 &= \left[ \ \ 1 \ \ \ \frac{\beta }{\alpha} \ \  0 \right]^\top \ \ \mbox{for} \ \alpha \neq 0 \ \  \text{or } \ \tilde{\vec v}_2 = \left[ \frac{\  \alpha}{\beta} \ \ 1 \ \ 0 \right]^\top \ \ \mbox{for} \ \beta \neq 0, \\
\noalign{\vskip5pt}
        \tilde{\vec v}_3 &= \left[ 0 \ \  0 \ \ 1 \right]^\top.
        \end{aligned}\label{eq:modes}
        \end{equation}
        Returning to the eigenvalue problem of interest, \eqref{eq:eigenvalueProblem2}, we simply need to scale the eigenvalues $\tilde{\lambda}_1, \tilde{\lambda}_2, \tilde{\lambda}_3$ by the multiplicative constant $K_{\alpha, \beta}/\mu$ in the continuous case, or $\Kd/\mu$ in the discrete case. For the latter case, the eigenvalues are

        \begin{equation}
            \lambda_1 = \frac{\Kd}{\mu} \tilde{\lambda}_1,\hspace{.2cm} \lambda_2 = \frac{\Kd}{\mu} \tilde{\lambda}_2 ,\hspace{.2cm}\lambda_3 = \frac{\Kd}{\mu} \tilde{\lambda}_3. \label{eq:lambdasKT}
        \end{equation}   
        
        The eigenvalues in \eqref{eq:lambdasKT} are negative for all the blob functions we consider, corresponding to a stable equilibrium configuration. This is verifiable by plotting the eigenvalues as functions of $\alpha,\beta$. From the form of the eigenvectors $\tilde{\vec v}_1,\tilde{\vec v}_2$ in \eqref{eq:modes}, we see that the associated eigenvalues $\tilde{\lambda}_1,\tilde{\lambda}_2$ in \eqref{eq:lambdas} are decay rates of perturbations to the surface that are parallel to the equilibrium configuration in the $z=0$ plane. The eigenvalues $\tilde{\lambda}_3$ in \eqref{eq:lambdas} correspond to the decay rates of perturbations orthogonal to this plane. 

        We note that in the zero wavenumber case, the regularized Stokeslet matrix \eqref{eq:Scoef0} has nonzero eigenvalues. However, since $K_{\alpha,\beta}$ and $\Kd$ vanish in the zero wavenumber case, the corresponding eigenvalues in the linearized surface problem will also vanish.

\section{Results and Discussion} 
\label{sec:results}
In this section, we highlight results that follow from the linear stability analysis in the previous section. We begin by focusing on the specific case of the doubly periodic regularized Stokeslet corresponding to the algebraic blob function in \eqref{eq:algblob}. We then compare these results to other blob functions and numerically verify our findings by estimating the critical time step for 
 a forward Euler method. We conclude with a numerical study of a finite elastic surface in a bulk fluid and show that the theoretical results for the doubly periodic domain are also relevant in this setting.

We will use typical values for the nondimensional tensile and bending stiffness for structures at the microscale, which generally are several orders of magnitude different from each other \cite{jabbarzadeh2020, olson2011,Yu2022}. For example, the nondimensional tensile and bending stiffness in a mammalian sperm model were chosen as $\sigma=75$ and $\kappa_B=0.075$, respectively  \cite{olson2011}. While these were used initially for a 1D Euler elastica model, we extend their use to our 2D elastic surface in this study and set $\sigma_s=\sigma_q=75$ and $\kappa_B=0.075$. 
In the cases where we use a surface with tensile rigidity only (no bending rigidity), we set $\kappa_B=0$.

In addition, we normalize the viscosity, $\mu=1$, and the length of the periodic square, $L=1$. The surface is discretized with $N=32$ points in each direction ($32^2=1024$ points in total) so that the distance between points in the equilibrium configuration is $h=1/32$. The choice of $N=32$ constrains the wavenumbers $\alpha,\beta \in \mathbb{F}_{32}$.

\subsection{Varying Regularization and Discretization Parameters} 
In the most common implementations of the MRS, the choice of regularization parameter $\epsilon$ and the discretization length $h$ is crucial to ensure accurate evaluations of the velocity field. Importantly, the accuracy of the method relies on choosing $\epsilon$ in tandem with the average spatial discretization of the boundary, $h$. As shown in \cite{cortez2005}, a good rule of thumb is to choose $\epsilon=\mathcal{O}(h)$. When $\epsilon \gg h$, the regularization error dominates, whereas when $\epsilon \ll h$, the discretization error dominates and the boundaries tend to allow fluid to leak through them.

\begin{figure}[htb!]
    \centering
    \begin{subfigure}[t]{0.5\textwidth} 
        \centering    
        \includegraphics[width=\textwidth]{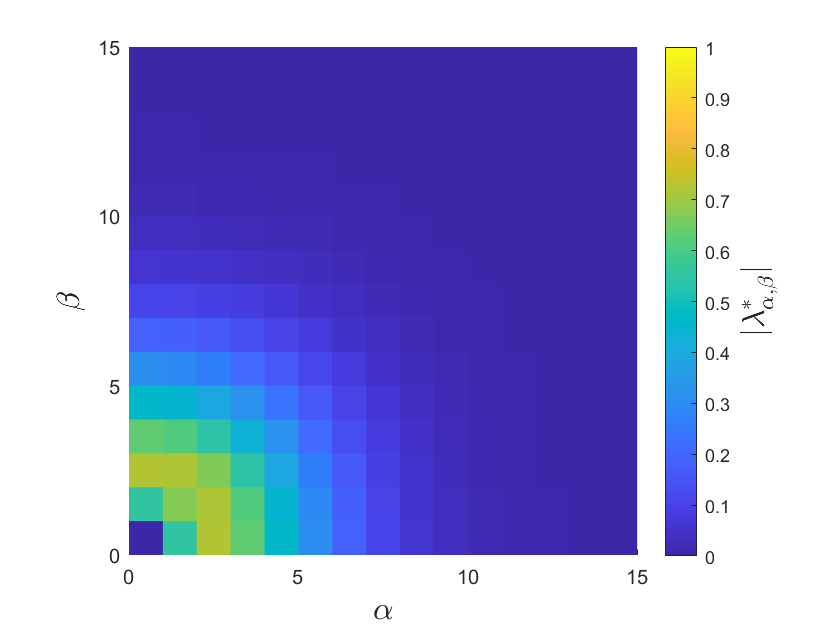}  
         \caption{$\epsilon/h=4$}
        \label{subfig:heatmap1}
    \end{subfigure}\begin{subfigure}[t]{0.5\textwidth} 
        \centering
        \includegraphics[width=\textwidth]{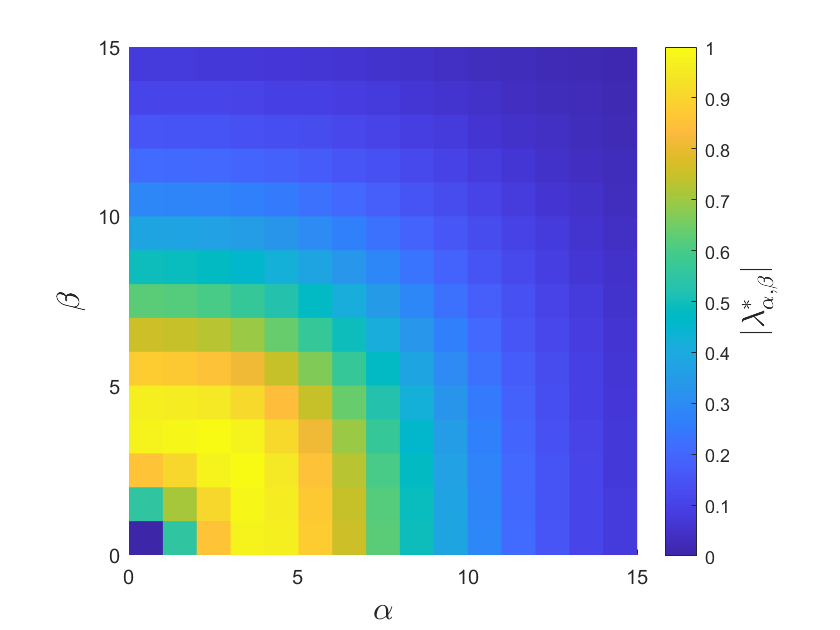}    
        \caption{$\epsilon/h=2$}
        \label{subfig:heatmap2}
    \end{subfigure}
    
    \begin{subfigure}[t]{0.5\textwidth} 
        \centering    
        \includegraphics[width=\textwidth]{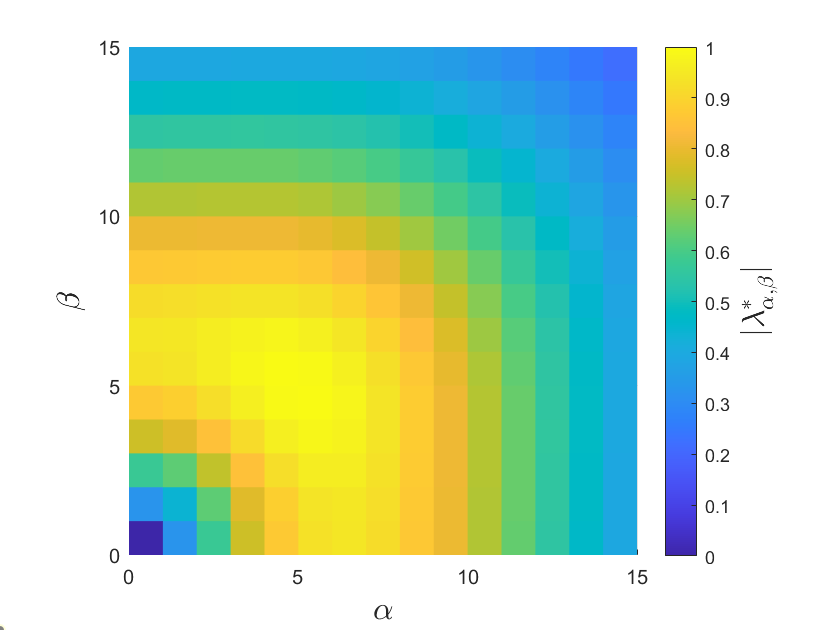}  
        \caption{$\epsilon/h=1$}
        \label{subfig:heatmap3}
    \end{subfigure}\begin{subfigure}[t]{0.5\textwidth} 
        \centering
        \includegraphics[width=\textwidth]{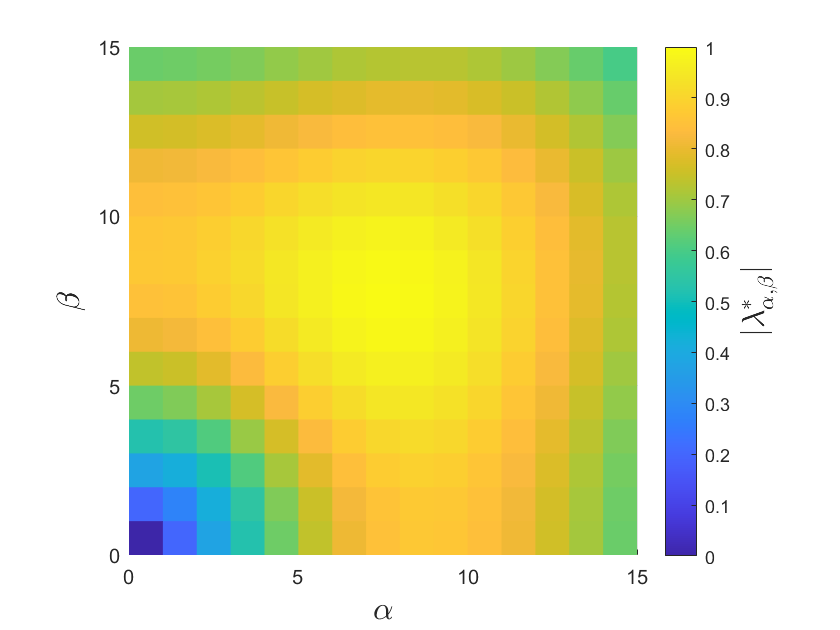}    
        \caption{$\epsilon/h=1/2$}
        \label{subfig:heatmap4}
    \end{subfigure}\caption{Scaled eigenvalues from \eqref{eq:eigneg} as a function of wavenumbers $\alpha, \beta$ for four different
scalings of $\epsilon/h$: (a) 4, (b) 2, (c) 1, and (d) 1/2. The eigenvalues are normalized by the most negative eigenvalue, $\min_{\alpha,\beta}(\lambda_1,\lambda_2,\lambda_3)$, for each $\epsilon/h$. These decrease as $\epsilon/h$ decreases and are -172
in (a), -344 in (b), -652 in (c), and -1118 in (d).}
    \label{fig:epsilonVaried}
\end{figure}

For the case of an elastic surface with tensile rigidity but no bending rigidity, we explore how the eigenvalues $\lambda_1,\lambda_2, \lambda_3$ from \eqref{eq:lambdasKT} vary for different scalings of $\epsilon/h$ when utilizing $\phi_\epsilon$ from \eqref{eq:algblob}.  Fig.~\ref{fig:epsilonVaried} is a heat map of the magnitude of the most negative scaled eigenvalues $\lambda^*_{(\alpha,\beta)}$ across all nonnegative wavenumbers $\alpha,\beta = 0, 1, \ldots 15$, defined as 
\begin{equation}
    \lambda^* = \lambda^*_{\alpha,\beta}= \frac{ \min \left ( \lambda_1(\alpha,\beta), \lambda_2(\alpha,\beta), \lambda_3(\alpha,\beta) \right )}{ \min_{\alpha,\beta} \left (\lambda_1, \lambda_2, \lambda_3 \right )}.\label{eq:eigneg}
\end{equation}The minimum in the numerator is taken for a fixed $\alpha,\beta$ whereas the minimum in the denominator is taken across the wavenumbers $\alpha,\beta = 0, 1, \dots, 15$. For relatively larger regularizations ($\epsilon/h = 4, 2$ in Fig.~\ref{fig:epsilonVaried}(a),(b)), we observe that the most negative eigenvalues (corresponding to the ones with the largest decay rates) appear for smaller wavenumbers. 
When the regularization is decreased so that $\epsilon/h \leq 1$, the most negative eigenvalues are distributed over a larger region containing the intermediate wavenumbers. Considering $\epsilon$ as a length scale which is inversely proportional to the wavenumber magnitude, $\sqrt{\alpha^2+\beta^2}$, we interpret this to mean that as the regularization parameter $\epsilon$ decreases compared to $h$, it is the smaller length scale dynamics that determine the time step necessary for stability. On the other hand, the smoothing provided by a larger $\epsilon/h$ attenuates the rapid changes at small length scales. 

\subsection{Numerical Stability Comparison for Different Blob Functions} 
In the following, we compare the theoretical results of the linear stability analysis for several blob functions. These results are then checked with the results of numerical simulations using the \textit{nonlinear} boundary integral equations \eqref{eq:nonlinearSystem}. 

We have already introduced the algebraic blob \eqref{eq:algblob}. In this section, we refer to this blob as $\phi^A_{\epsilon}(r)$ where the superscript $A$ denotes ``algebraic." Now we consider another algebraic blob,

\begin{equation}
    \label{eq:algBlob2}
    \phi^{A,3M}_{\epsilon}(r) = \frac{15 \epsilon^4 \left ( 40 \epsilon^6 -132 \epsilon^4r^2+57\epsilon^2r^4 - 2r^6 \right )}{16 \pi \left ( r^2 + \epsilon^2 \right )^{13/2}},
\end{equation}where the superscript $3M$ denotes the total number of moment conditions this blob satisfies. These moment conditions improve the convergence of the far-field regularization error and are as follows:

\begin{equation}
    \label{eq:secondMoment}
    \int_0^{\infty}\ \phi_{\epsilon}(s)s^4\ ds = 0
\end{equation}and

\begin{equation}
    \label{eq:momentBlob}
    \int_0^{\infty} \phi_{\epsilon}(s) s^{2k+1} \ ds = 0, \ k=1,2.
\end{equation}Equation \eqref{eq:secondMoment} is often referred to as the second moment condition and ensures that the far-field velocity error (compared to the singular Stokeslet) decays faster than $O(\epsilon^2)$ \cite{nguyen2014, zhao2019}. The two additional conditions in \eqref{eq:momentBlob} improve the regularization error specifically in the case of surface forces \cite{beale2001, nguyen2025}. For comparison, $\phi_\epsilon^A$ in \eqref{eq:algblob} does not satisfy any moment conditions. 

\begin{figure}[htb!]
    \centering
    \includegraphics[width=0.9\textwidth]{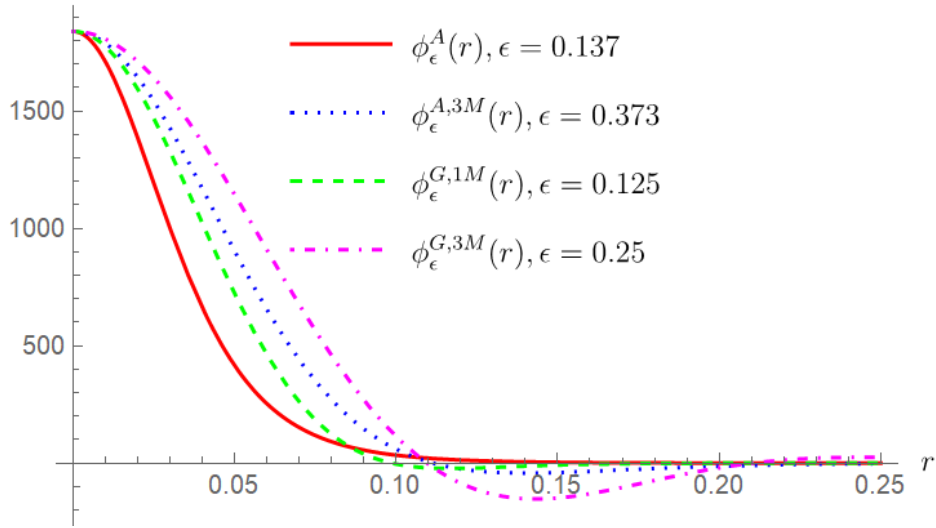}
    \caption{Different blob functions as a function of the radial distance $r$. The regularization parameters $\epsilon$ are chosen so that the maximum of each is the same. Note that because of the second-moment condition \eqref{eq:secondMoment}, every blob function besides $\phi^A_{\epsilon}$ must have positive and negative sections.}
    \label{fig:blobsComparison}
\end{figure}

Additionally, we consider a pair of Gaussian-type blob functions,

\begin{equation}
    \label{eq:gaussianBlob}
    \phi^{G,1M}_{\epsilon}(r) = \frac{ \exp \left ( -r^2 / \epsilon^2 \right ) }{\pi^{3/2} \epsilon^3} \left ( 5/2 - r^2/\epsilon^2 \right )
\end{equation}and

\begin{equation}
    \label{eq:gaussianBlob2}
    \phi^{G,3M}_{\epsilon}(r) = \frac{2 \exp \left ( -r^2 / \epsilon^2 \right ) }{3 \pi^{3/2} \epsilon^9} \left ( -2r^6 + 21r^4 \epsilon^2 - 54 r^2 \epsilon^4 + 30 \epsilon^6 \right ).
\end{equation}The blob $\phi^{G,1M}_{\epsilon}(r)$ satisfies only the second moment condition \eqref{eq:secondMoment}, while the blob $\phi^{G,3M}_{\epsilon}(r)$ satisfies both \eqref{eq:secondMoment} and \eqref{eq:momentBlob}. In Fig.~\ref{fig:blobsComparison}, we plot all four blobs as functions of the radial distance $r$, choosing an $\epsilon$ that ensures the same peak value when $r=0$. The formulas of the doubly periodic regularized Stokeslets associated with these blobs are given in \ref{app:blobmoment} and details about their numerical implementation are given in \ref{app:ffts}.

We consider the numerical stability of the elastic surface problem for the different blob functions using forward Euler time integration. Although other time integration methods may have advantages in accuracy or stability, forward Euler remains the most commonly used in MRS simulations due to its relative simplicity. Indeed, fully implicit implementations require solving a large, dense system of nonlinear equations at every time step. Other explicit methods, such as classical higher-order Runge-Kutta (RK) methods, are sometimes utilized (e.g. RK4 in \cite{Yu2022}). However, since the eigenvalues for this particular linearized problem are real and negative, in terms of stability, it is not ``more efficient" to use higher-order classical RK methods. In this sense, we mean that if one compares the absolute stability regions of these methods, forward Euler (a.k.a. RK1) covers a longer segment of the negative real line per stage than the others. We note that the family of Runge-Kutta-Chebyshev methods, given their extended region of stability over the negative real line, may offer an improvement in stability for this problem \cite{ascher2008}; however, for simplicity, we will focus here on forward Euler time integration. 

\begin{figure*}[htb!]
    \centering
    \begin{subfigure}[t]{0.5\textwidth} 
        \centering          \includegraphics[width=\textwidth]{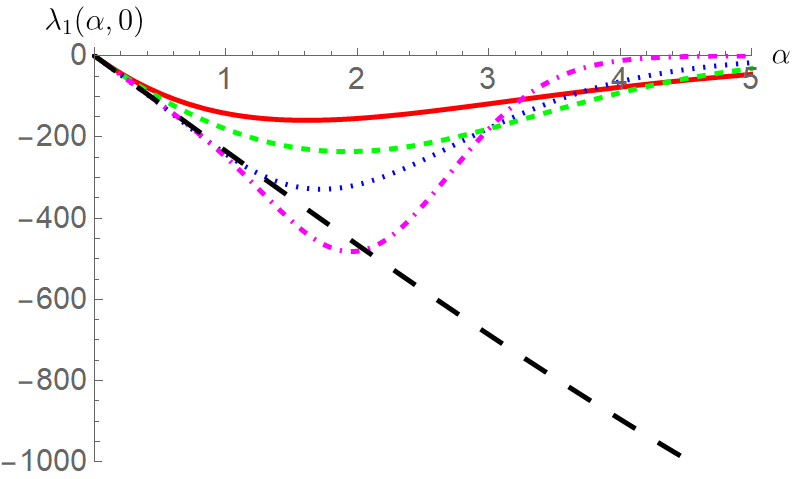}  
        \caption{Eigenvalue $\lambda_1$}
        \label{subfig:lambda1}
    \end{subfigure}\begin{subfigure}[t]{0.5\textwidth} 
        \centering
        \includegraphics[width=\textwidth]{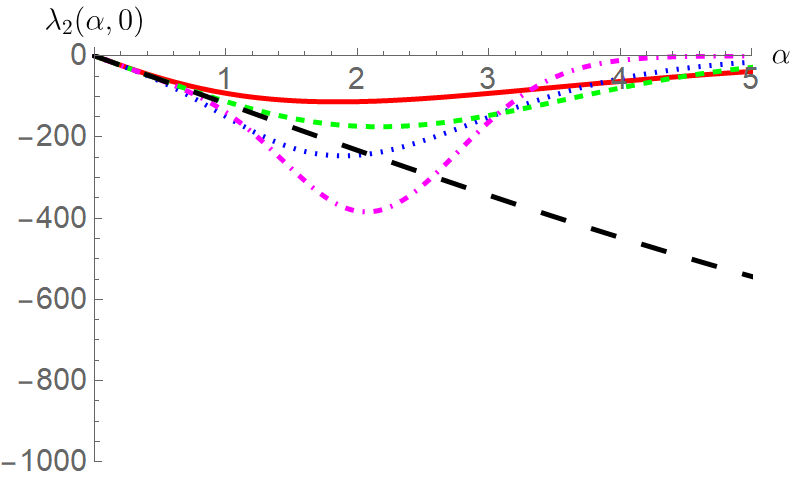}    
        \caption{Eigenvalue $\lambda_2$}
    \end{subfigure}
    
    \begin{subfigure}[t]{1.0\textwidth} 
        \centering
         \includegraphics[width=0.2\textwidth]{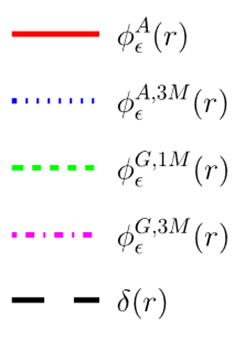}
        \includegraphics[width=0.5\textwidth]{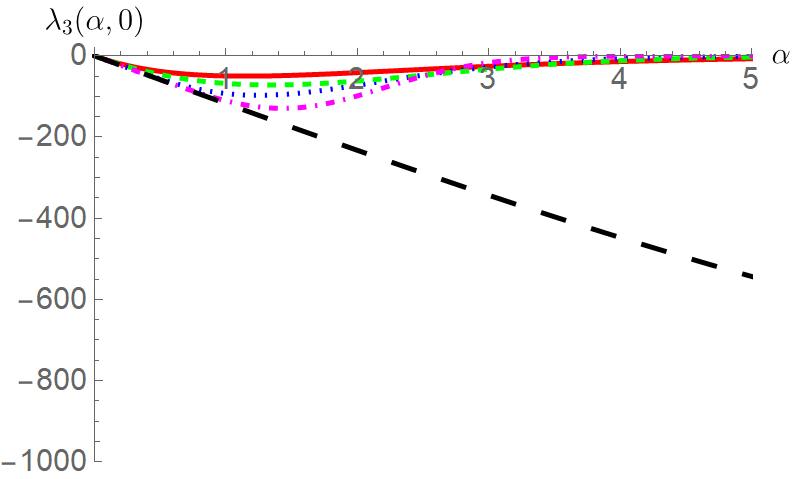}    
        \caption{Eigenvalue $\lambda_3$}
    \end{subfigure}  
    \caption{Eigenvalues $\lambda_1$, $\lambda_2$, and $\lambda_3$ in (a)-(c), respectively, with $(\alpha, \beta)=(\alpha,0)$. The different blob functions are specified in the legend and the same as those shown in Fig.~\ref{fig:blobsComparison}. The eigenvalues for the associated singular Stokeslet (denoted by $\delta$) are shown for comparison (given in  \cite{pozrikidis1996}).}
    \label{fig:eigenvalues}
\end{figure*}

For a given blob function, $\phi_{\epsilon}$, the theoretical critical time step $t_c$ for our elastic surface-fluid system discretized in time by forward Euler is

\begin{equation}
    t_c = \left | 2 / \min_{\alpha,\beta} \left (\lambda_1, \lambda_2, \lambda_3 \right ) \right |
\end{equation}where $\min_{\alpha,\beta} \left (\lambda_1, \lambda_2, \lambda_3 \right )$ is the most negative eigenvalue, and $\lambda_1, \lambda_2, \lambda_3$ are as defined in \eqref{eq:lambdasKT}. In addition to the choice of blob function and regularization parameter $\epsilon$, each eigenvalue depends on the choice of elastic model parameters. Here, we use a surface with tensile rigidity only so that $\kappa_B=0$. We plot the eigenvalues $\lambda_1, \ \lambda_2, \ \lambda_3$  for a range of $\alpha$ and $\beta = 0$, in Fig.~\ref{fig:eigenvalues}. For each blob, we use the regularizations shown in Fig.~\ref{fig:blobsComparison} so that the maximum of each blob is the same, allowing for a fair comparison. 

From Fig.~\ref{fig:eigenvalues}, we observe that the most negative eigenvalue for all blob functions is $\lambda_1$. Recall from \eqref{eq:modes} that this is one of the eigenvalues associated with motion of the surface tangential to the $z=0$ plane. The other eigenvalue associated with this tangential motion is $\lambda_2$, while $\lambda_3$ corresponds to motion orthogonal to this plane. For all blobs, $|\lambda_1|$ is greater than twice $|\lambda_3|$, revealing that the maximum time step constraint is due to motion tangential to the plane. This pattern continues for other sets of wavenumbers (e.g. $\beta \neq 0$).

Comparing results between the different blobs, we observe that the smallest nonzero discrete Fourier mode ($\alpha = 1$) is captured best by the blobs $\phi_{\epsilon}^{G,3M}$ and $\phi_{\epsilon}^{A,3M}$, i.e. the ones that satisfy all of the moment conditions considered. The price of this better agreement with the singular Stokeslet is a more stringent time step constraint. By comparing the minima of the eigenvalues in Fig.~\ref{subfig:lambda1}, we observe that the time step constraint for simulations using $\phi_{\epsilon}^{G,3M}$ will be less than half of the constraint imposed when using $\phi_{\epsilon}^A$. 

To validate the theoretical results of the spatially discretized linear analysis, we numerically find the time step at which the system becomes unstable for each nonlinear simulation. We take the same approach of \cite{huapeskin} in finding this critical time step. At each time step, we measure the discrete $\ell_2$ norm of the surface velocity. We simulate up to $T_{\text{final}}=10$. If the $\ell_2$ norm of the velocity increases over ten consecutive time steps, we consider the time step unstable and use a smaller one. By using a binary search, we can determine the numerical critical time step $t_c$ up to arbitrary precision.

In the tests, we initialize the surface with a small sinusoidal perturbation to the equilibrium configuration

\begin{equation}
    \vec X(q_j,s_k) = \begin{bmatrix} q_j \\ s_k \\ 0 \end{bmatrix} + 0.01 \begin{bmatrix} 0 \\ \sin \left (4 \pi q_j \right )  \\ 
\sin \left (2 \pi (q_j + s_k) \right ) + \cos \left ( 4 \pi q_j \right ) \end{bmatrix} 
\label{eq:perturbation}
\end{equation}
where $(q_j,s_k) = (j/32, k/32)$ for $j,k\in\mathbb{Z}_{32}$. In the simulations, we utilize an Inverse Fast Fourier Transform (IFFT) to evaluate the velocity of the elastic surface at each time step. This is based on the method of \cite{hoffmann1} and the basic idea is summarized in Algorithm \ref{alg:fft} of \ref{app:ffts}. Additional details on its implementation are also discussed in \ref{app:ffts}. Two example videos from these simulations demonstrating stable and unstable behavior are available in \ref{app:videos}.

    \begin{table}[ht]
        \centering 
        \resizebox{\textwidth}{!}{
        \begin{tabular}{l|c|c|c||c|c|c}
        & \multicolumn{3}{c}{$\phi^{G,1M}_{\epsilon}$} & \multicolumn{3}{c}{$\phi^{G,3M}_{\epsilon}$}  \\
        
        & \bfseries $\epsilon$ & \bfseries The. $t_c$ &\bfseries Num. $t_c$ &  \bfseries $\epsilon$ & \bfseries The. $t_c$ &\bfseries Num. $t_c$ \\
        
        & 0.125 & $\mathbf{8.499413}1e{-3}$ & $\mathbf{8.499413}0e{-3}$  & 0.25 & $\mathbf{4.1578}719e{-3}$ & $\mathbf{4.1578}979e{-3}$ \\

         (T) & 0.0625 & $\mathbf{4.316093}31e{-3}$ & $\mathbf{4.316093}09e{-3}$ & 0.125 & $\mathbf{2.14419}14{-3}$ & $ \mathbf{2.14419}07e{-3}$ \\ 

        & 0.03125 & $\mathbf{2.29}12115e{-3}$ &
         $\mathbf{2.29}03107e{-3}$ & 0.0625 & $\mathbf{1.134912}4{-3}$ & $\mathbf{1.134912}1e{-3}$ \\

        \hline \hline

         & 0.125 & $\mathbf{7.24079}21e{-3}$ & $\mathbf{7.24079}51e{-3}$  & 0.25 & $\mathbf{3.59710}26e{-3}$ & $\mathbf{3.59710}32e{-3}$ \\ 
         (TB) & 0.0625 & $\mathbf{2.404681}3e{-3}$ & $\mathbf{2.404681}2e{-3}$ & 0.125 & $\mathbf{1.2773098}e-{3}$ & $ \mathbf{1.2773098}e-{3}$ \\ 
        
         & 0.03125 & $\mathbf{6.0}688377e{-4}$ & $\mathbf{6.0}554777e{-4}$ & 0.0625 & $\mathbf{3}.3409256e{-4}$ & $ \mathbf{3}.251250000e{-4}$ \\
        \hline
        \end{tabular}}
        \caption{Theoretical  (The.) and Numerical (Num.) critical time steps $t_c$ for Gaussian blobs $\phi^{G,1M}_{\epsilon}$ and $\phi^{G,3M}_{\epsilon}$. (T) corresponds to simulations of an elastic surface with tensile rigidity only and (TB) to an elastic surface with tensile and bending rigidity.}
        \label{tab:gaussiantables}
    \end{table}

    \begin{table}[ht]
        \centering 
        \resizebox{\textwidth}{!}{
        \begin{tabular}{l|c|c|c||c|c|c}
        & \multicolumn{3}{c}{$\phi^{A}_{\epsilon}$} & \multicolumn{3}{c}{$\phi^{A,3M}_{\epsilon}$}  \\
        
        & \bfseries $\epsilon$ & \bfseries The. $t_c$ &\bfseries Num. $t_c$ &  \bfseries $\epsilon$ & \bfseries The. $t_c$ &\bfseries Num. $t_c$ \\
        
        & 0.137 & $\mathbf{1.26793}71e{-2}$ & $  \mathbf{1.26793}85e{-2}$ & 0.373 & $\mathbf{6.3033344}e-{3}$ & $\mathbf{6.3033344} e-{3}$ \\

         (T) & 0.0685 & $\mathbf{6.401184}0e{-3}$ & $ \mathbf{6.401184}1e{-3}$ & 0.1865 & $\mathbf{3.0956895}e-{3}$ & $\mathbf{3.0956895}e-{3}$ \\ 

        & 0.03425 & $\mathbf{3.3272270}e{-3}$ &
         $\mathbf{3.3272270}e{-3}$ & 0.09325 & $\mathbf{1.6316191}e{-3}$ & $\mathbf{1.6316191}e-{3}$ \\

        \hline \hline

         & 0.137 & $\mathbf{1.1225}511e{-2}$ & $\mathbf{1.1225}444e-{2}$ & 0.373 & $ \mathbf{5.4532080}e-{3}$ & $\mathbf{5.4532080}e-{3}$ \\
         
         (TB) & 0.0685 & $\mathbf{3.7197432}e-{3}$ & $\mathbf{3.7197432}e{-3}$ & 0.1865 & $\mathbf{1.9597904}e-{3}$ & $\mathbf{1.9597904}e-{3}$\\ 
        
         & 0.03425 & $\mathbf{9.4055}556e{-4}$ & $\mathbf{9.4055}396e{-4}$ & 0.09325 & $\mathbf{5.0885}336e{-4}$ & $ \mathbf{5.0885}288e-{4}$ \\
         
        \hline
        \end{tabular}}
        \caption{Theoretical  (The.) and Numerical (Num.) critical time steps $t_c$ for algebraic blobs $\phi^{A}_{\epsilon}$ and $\phi^{A,3M}_{\epsilon}$. (T) corresponds to simulations of an elastic surface with tensile rigidity only and (TB) to an elastic surface with tensile and bending rigidity.}
        \label{tab:algebraictables}
    \end{table}Results from the numerical simulations are recorded in Tables \ref{tab:gaussiantables} and \ref{tab:algebraictables} for the Gaussian and algebraic blobs, respectively. The theoretical critical time step prediction from the analysis is shown for comparison. The first three rows in each table correspond to the results for an elastic surface with tensile rigidity only (T). The last three rows of each table correspond to the results for an elastic surface with tensile and bending rigidity (TB).

    By looking at successive rows for a particular elastic surface model, we observe the effect on stability from halving the regularization with the fixed discretization $h=1/32=0.03125$. The largest regularizations tested correspond to those shown in Fig.~\ref{fig:blobsComparison}, so corresponding entries in the same row of either table correspond to blob functions with the same maximum. For example, using $\epsilon=0.125$ for $\phi^{G_{3M}}_{\epsilon}$ results in a blob function with the same peak as $\epsilon = 0.0685$ for $\phi^A_{\epsilon}$. 

    Results for both tables show strong agreement between the theoretical critical time step and the numerical one, verifying the analytical results. The agreement in significant digits is highlighted in bold in both tables. For the Gaussian blobs, we observe a loss in the estimation's precision for the smallest regularization tested in the (TB) cases. For the algebraic blobs, we see a loss in precision but it is less severe; the estimate still agrees up to five significant digits. We theorize that in the (TB) simulations, the effects of nonlinearity become more significant as $\epsilon$ is decreased, especially for the Gaussian blobs. A similar sized loss of precision for the algebraic blobs in both the (TB) models would likely occur as $\epsilon$ is decreased to a point where the linear approximation is less accurate.

    Comparing the time constraints between the (T) and (TB) models, we observe that the effect of bending rigidity on the stability becomes more significant as $\epsilon$ is decreased. For example, the numerical time step constraint for the (T) model with $\phi^A_{\epsilon}$ and $\epsilon=0.137$ is $t_c \approx 1.2e-2$ while the time step constraint for the (TB) model with the same parameters is $t_c \approx 1.1e-2$, approximately a decrease of only $10\%$. On the other hand, if we change the regularization to $0.03425$, the time step constraint for the (T) model is $t_c \approx 3.3e-3 $ whereas the constraint for the (TB) model is $t_c \approx 9.4e-4$, approximately a $70\%$ difference.

    \subsection{Numerical Tests with Finite surface}
    We conclude with a comparison of our theoretical results for the infinite, doubly periodic surface and numerical results of the same problem using a finite surface in a bulk fluid (infinite fluid with far field conditions, $\vec u\to 0$ as $|\vec x|\to \infty$). This is often a natural setting as finite-length micro-structures cannot necessarily be modeled as a periodic structure. 

    The finite surface with $x,y\in[0,L]\times[0,L]$ requires a modification of the tensile forces. Instead of \eqref{eq:tensionForce}, we use Hookean springs with a nonzero resting length so that the tension in the $q$ or $s$ material direction is now

    \begin{equation}
        T^q = \sigma_q \left ( \left | \pp{\vec X}{q} \right | - 1 \right ), \ \  T^s = \sigma_s \left ( \left | \pp{\vec X}{s} \right | - 1 \right ).
    \end{equation}The reason for this is that when we use springs with zero resting length for a finite sheet, the sheet will collapse in on itself. In the case of an infinite, doubly periodic sheet, every point on the sheet has a neighbor in either direction which prevents this from happening. The new tensile force is

    \begin{equation}
        \vec F_T(\vec X(q,s,t) ) = \pp{}{q} \left ( T^q \vecg {\tau^q} \right ) + \pp{}{s} \left ( T^s \vecg {\tau^s} \right )
        \label{eq:newTension}
    \end{equation}where $\vecg \tau^q$ and $\vecg \tau^s$ are the unit tangent vectors in the $q,s$ directions respectively. The discretized form of the tensile force \eqref{eq:newTension} for the surface with uniform resting length $h$ between neighboring points is

    \begin{equation}
        \begin{split}
        \vec F_{T_{j,k}} = \   &\sigma_q D_q \left [ \left ( \frac{\left |\vec X (q_j, s_k, t) - \vec X(q_{j-1}, s_k, t) \right | }{h} - 1  \right ) \vecg \tau^q_{j,k}\right ] \\
        + &\sigma_s D_s \left [ \left ( \frac{\left |\vec X (q_j, s_k, t) - \vec X(q_j, s_{k-1}, t) \right | }{h} - 1  \right ) \vecg \tau^s_{j,k}\right ]
        \label{eq:newTensionDiscrete}
        \end{split}
    \end{equation}where we have used the discrete tangential vectors in the $q, \ s$ directions

    \[ \vecg \tau^q_{j,k} = \frac{\vec X (q_j, s_k, t) - \vec X (q_{j-1}, s_k, t)}{\left | \vec X (q_j, s_k, t) - \vec X(q_{j-1}, s_k, t)\right |}, \ \ \vecg \tau^s_{j,k} = \frac{\vec X (q_j, s_k, t) - \vec X (q_j, s_{k-1}, t)}{\left | \vec X (q_j, s_k, t) - \vec X(q_j, s_{k-1}, t)\right |}.\]The discretized bending force $\vec F_{B_{j,k}}$ remains the same as in \eqref{eq:fbdiscrete}.  

    Again we will consider the case of a surface with only tensile rigidity, and the case of a surface with tensile and bending rigidity. In the latter case, the bending rigidity pushes the surface toward the flat configuration when out-of-plane deflections are initialized. But in the case of a surface with only tensile rigidity, we need to supply external forces at the edges to push it toward the flat configuration. These forces are 

    \begin{equation} 
    \begin{split}
        \vec F^{-x}_{j,k} = \left [-75 \ \ 0 \ \ 0 \right ]^{\top}&, \ \  \vec F^{+x}_{j,k} = \left [75 \ \ 0 \ \ 0 \right ]^{\top} \\ 
         \vec F^{-y}_{j,k} = \left [0 \ \ -75 \ \ 0 \right ]^{\top}&, \ \
        \vec F^{+y}_{j,k} = \left [0 \ \ 75 \ \ 0 \right ]^{\top} \ \   
        \end{split}
    \end{equation}where the $\mp x$ superscripts correspond to the edges of the surface at $q=0,L$, and the $\mp y$ superscripts correspond to the edges of the surface at $s=0,L$, both in the reference configuration. These external forces effectively stretch the surface at its edges, which due to the internal tensile forces, pushes it toward a flat configuration. 

       \begin{figure}[htb!]
        \centering   
         \begin{subfigure}[t]{0.5\textwidth}
            \includegraphics[width=1.0\textwidth]{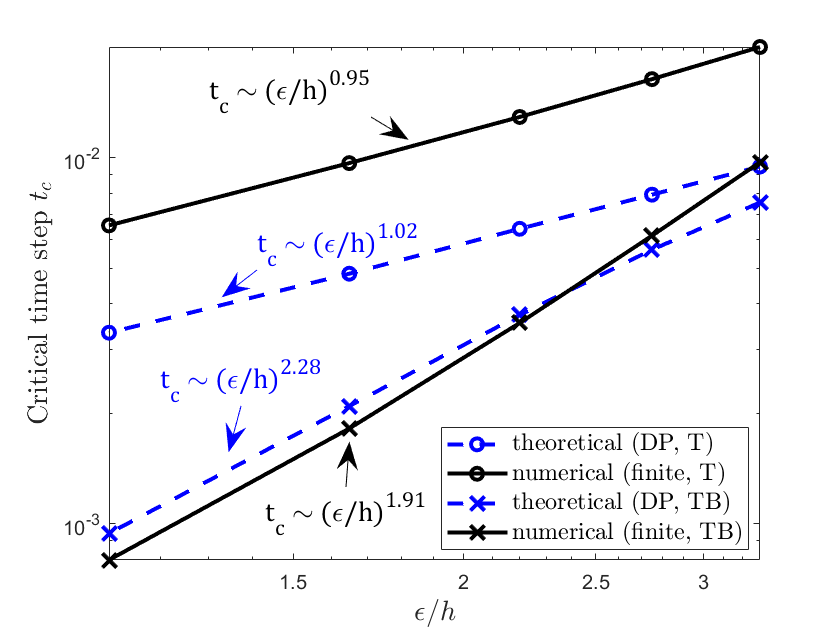}
        \caption{Surface discretization $h=1/32$}
        \label{fig:finiteN32}
        \end{subfigure}\begin{subfigure}[t]{0.5\textwidth}
            \includegraphics[width=1.0\textwidth]{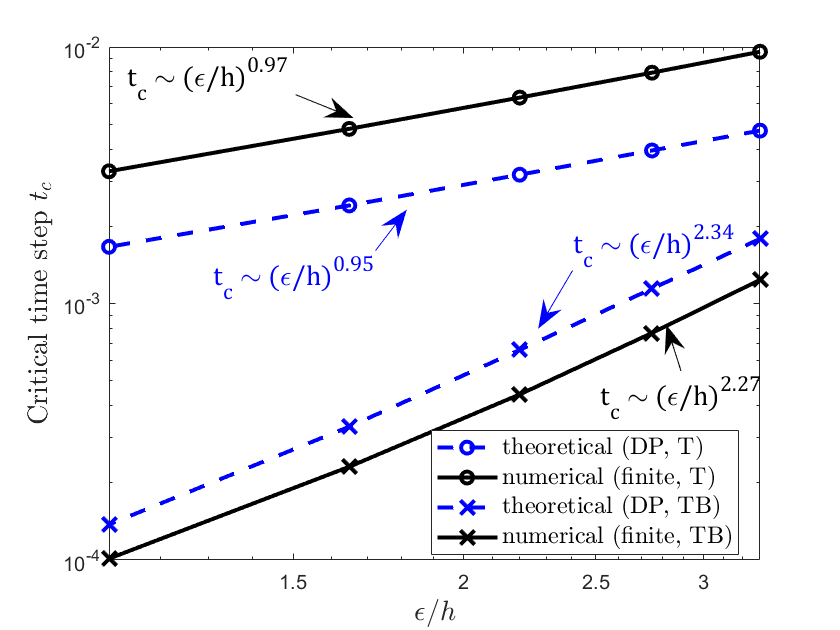}
        \caption{Surface discretization $h=1/64$}
        \label{fig:finiteN64}
        \end{subfigure}
        \caption{Log-log plots for the empirical critical time step $t_c$ for a finite surface in black and the theoretical critical time step from the doubly periodic Fourier analysis in blue. (T) denotes an elastic surface with tensile rigidity only ($\sigma_q=\sigma_s=75$) and (TB) denotes an elastic surface with the same tensile rigidity and bending rigidity $\kappa_B=0.075$.}
        \label{fig:finite}
    \end{figure}
    
     In this finite surface test, we focus our attention on a single blob function, $\phi^A_{\epsilon}$. We test this for a surface discretized with $N=32,64$ points in each direction so that $h=1/32, \ 1/64$. Since we are not in the doubly periodic setting, we use the classic Nystr{\"o}m implementation of the MRS with a trapezoidal quadrature. The surface is initialized with the sinusoidal perturbation in \eqref{eq:perturbation}. An example video of the simulation is available in \ref{app:videos}. As before, we do a binary search for the forward Euler critical time step $t_c$ by measuring the discrete $\ell_2$ norm of the surface velocity. We completed this search for $\epsilon/h = \left(\frac{3 \sqrt \pi}{4}\right)^{1/3} \{3, \ 5/2, \ 2, \ 3/2, \ 1\}$ (the factor $\left(\frac{3 \sqrt \pi}{4}\right)^{1/3} \approx 1.1$ is the same factor used to make the peak of $\phi_{\epsilon}^A$ the same height as the peak of $\phi_{\epsilon}^{G,1M}$).

     In Fig.~\ref{fig:finite}, we compare the numerical critical time steps for the finite surface with the theoretical ones predicted by the analysis in the doubly periodic domain. The log-log plots show that both the theoretical and numerical critical time steps are approximated well by a power law in terms of $\epsilon/h$, the ratio of the regularization parameter to the spatial discretization of the surface. These power laws are found by minimizing errors in a least squares sense. For the (T) model (surface with only tensile rigidity), we observe that the numerical critical time step is approximated well by a slightly sublinear function of $\epsilon/h$ ($t_c \sim (\epsilon/h)^{0.95}$ for $h=1/32$ and $t_c \sim (\epsilon/h)^{0.97}$ for $h=1/64$). This approximation agrees well with the theoretical estimate for $h=1/64$, where the relationship is approximated by $t_c \sim (\epsilon/h)^{0.95}$. 

     For the (TB) model (surface with tensile and bending rigidity), we observe that the numerical critical time step is approximated well by a nearly quadratic function of $\epsilon/h$. For the case of $h=1/32$ we have $t_c \sim (\epsilon/h)^{2.28}$, and for the case of $h=1/64$ we have $t_c \sim (\epsilon/h)^{2.34}$. These agree well with the theoretically predicted relationship for $h=1/64$ that says $t_c \sim (\epsilon/h)^{2.27}$. 
     
     We note that while the power law relationship from the theory with $h=1/64$ agrees well with the empirical relationships, it is not as accurate for $h=1/32$. We expect that this is due to the effect of periodicity in this setting. As the grid size becomes larger, the regularization must also increase. But since $\epsilon$ is a parameter that is proportional to the width of the blob function, the contribution from periodic copies of forces becomes less negligible. By decreasing the discretization $h$, the regularization parameter is also decreased so that periodic effects are less relevant.

 \section{Conclusion}
\label{sec:conclusion}
A linear stability analysis for a coupled elastic surface-fluid system formulated with the MRS has been presented. The analytical results were validated numerically for several different blob functions. We found that the main constraint on the time step was due to the tangential motion of the 2D elastic surface and that the associated eigenvalue ($\lambda_1$ in Eq. \eqref{eq:lambdasKT}) is proportional to the regularized doubly periodic Green's function, $\widehat{G}^{DP}_{\phi_\epsilon}$. To our knowledge, the formulae for the doubly periodic Stokeslets associated with the Gaussian blob function $\phi_\epsilon^{G,3M}$ and the algebraic blob functions, $\phi_{\epsilon}^{A}$ and $\phi_{\epsilon}^{A,3M}$, have not appeared in the literature until now. Their derivation allowed us to show that when using blob functions that satisfy moment conditions like $\eqref{eq:secondMoment}$ and $\eqref{eq:momentBlob}$, better agreement with the singular Stokeslet comes at the cost of a more stringent time step to preserve numerical stability. 

The last part of Section \ref{sec:results} demonstrated the relevance of the stability analysis in the case of a finite surface in a bulk fluid with typical nondimensional values for tensile/bending rigidity. For the algebraic blob $\phi_{\epsilon}^A$, we showed that the empirical time step constraint for a surface with only tensile rigidity scaled almost linearly with $\epsilon/h$. In the case of a surface with tensile and bending rigidity as well, the time step constraint was greater than quadratic in $\epsilon/h$, indicating the strong dependence of stability imposed by the choice of regularization parameter $\epsilon$ for a fixed discretization $h$ when simulating elastic structures that resist bending. The empirical relationships from the finite surface case were shown to agree well with the theoretical relationships for a doubly periodic surface with a finer discretization. Overall, this work provides guidance for the practitioner using the MRS on the choice of time step based on the choice of blob function, regularization parameter, and elastic model.

\appendix

\section{The Method of Regularized Stokeslets}\label{app:regstokes}
\renewcommand{\theequation}{A.\arabic{equation}} 
\setcounter{equation}{0}
The method of regularized Stokeslets (MRS) is based on solutions to the Stokes equations \eqref{eq:st} and  \eqref{eq:div}, with a regularized forcing $\vec f = \vec g \phi_{\epsilon}(r)$, where $\vec g$ is a constant force, $r= |\vec r| = |\vec x - \vec y|$ is the displacement from the location of the force $\vec y$ to the observation point $\vec x$, and $\phi_{\epsilon}(r)$ is a mollifier, commonly referred to as a blob function. The blob function approximates the Dirac delta function $\delta(r)$ and satisfies $\iiint_\mathbb{R}^3 \phi_{\epsilon}(r) \ dV(\vec r)=1$ and $\lim_{\epsilon \to 0} \phi_{\epsilon}(r) = \delta(r)$.  

A solution to the Stokes equations with regularized forcing can be constructed through the functions $G_{\epsilon}(r)$ and $B_{\epsilon}(r)$, which satisfy $\Delta G_\epsilon(r) = \phi_{\epsilon}(r)$ and $\Delta B_{\epsilon}(r)= G_{\epsilon}(r)$. These can be thought of as regularized Green's functions for Laplace's equation and the biharmonic equation, respectively. 

We use these Green's functions to construct the regularized Stokeslet, $\Stokes$, which allows us to write the solution to the Stokes equations with regularized forcing as $\vec u(\vec x) = \frac{1}{8 \pi \mu} \Stokes(\vec r) \cdot \vec g$. In component form, the regularized Stokeslet can be written 

\[\mathcal{S}_{\phi_{\epsilon},ij} = 8 \pi \left [- G_{\phi_{\epsilon}}(r) \delta_{ij} + \pp{}{r_i} \pp{}{r_j} B_\epsilon(r) \right ]\]where $\delta_{ij}$ is the Kronecker delta and the partial derivatives are taken with respect to the components of the vector $\vec r = \vec x - \vec y$. 

In applications, forces are distributed over structures which may be modeled as 1D curves or 2D surfaces. Typically, we have a Lagrangian description of the structure $\vec X=\vec X(\vec q)$ where $\vec q$ are material coordinates in a reference configuration. We represent solutions as a convolution between the regularized Stokeslet kernel and the force density over the surface $\vec F(\vec q)$, 

\[ \vec u(\vec x;t) = \frac {1}{8 \pi \mu} \int_{\Gamma_t} \Stokes \left ( \vec x - \vec X(\vec q,t) \right )\cdot \vec F(\vec X(\vec q,t)) \ dS(\vec q). \]In practice, the structure is discretized and the integral is approximated by a quadrature method, 

\[ \vec u(\vec x; t) \approx \frac{1}{8 \pi \mu} \sum_{k=1}^N \ \Stokes( \vec x - \vec X(\vec q_k,t) ) \cdot \vec F(\vec X(\vec q_k)) \ w(\vec q_k), \]where $\vec q_k$ are the quadrature points, $k=1, \dots, N$, and $w(\vec q_k)$ are the weights associated with the points.

 \section{Doubly-periodic Regularized Stokeslet for the Blob Function $\phi_\epsilon=\phi_{\epsilon}^A$}\label{app:DPSB} 
\renewcommand{\theequation}{B.\arabic{equation}}
\setcounter{equation}{0}

As described in Section \ref{sec:blobChoice}, to determine the doubly-periodic regularized Stokeslet $\boldsymbol{\mathcal{S}}_{\phi_\epsilon}^{DP}$ and Fourier coefficients $\widehat{\boldsymbol{\mathcal{S}}}_{\phi_\epsilon}^{DP}$, we first need to choose a specific blob function $\phi_\epsilon$ and define its doubly-periodic extension. We will describe this in detail for the commonly used 3D algebraic blob function given in \eqref{eq:algblob} \cite{cortez2001}. Other doubly-periodic regularized Stokeslets are derived similarly and are recorded in \ref{app:blobmoment}. 

 The most commonly used blob functions in the MRS are radially symmetric, $\phi_{\epsilon}=\phi_{\epsilon}(r)$, resulting in a 2D Fourier transform that is radially symmetric in terms of the Fourier variables $\alpha, \beta$. That is, defining $k=\frac{2\pi}{L} \sqrt{\alpha^2+\beta^2}$, we have $\widehat{\phi}_{\epsilon}=\widehat{\phi}_{\epsilon}(z,k)$. Given this symmetry, we can evaluate the 2D Fourier transform $\phi_{\epsilon}$ as

\begin{equation}
    \widehat{\phi}_{\epsilon}^{DP}(z,k) = 2\pi \int_0^{\infty} J_0(k\rho) \phi_{\epsilon}(\rho,z) \rho \ d\rho \label{eq:blobJ}
    \end{equation}
    where $\rho = \sqrt{x^2+y^2}$ and $J_0$ is the zeroth order Bessel function of the first kind.
    
Specifically, for the algebraic blob in \eqref{eq:algblob}, we evaluated \eqref{eq:blobJ} with the aid of Mathematica and obtained

\begin{equation}
    \widehat{\phi}_\epsilon^{DP}(z,k)= \frac{\exp \left (-k \sqrt{z^2+\epsilon^2} \right ) \epsilon^4}{4} \left [ \frac{ 3 + 3k \sqrt{z^2+\epsilon^2} }{(z^2+\epsilon^2)^{5/2}} + \frac{k^2}{(z^2+\epsilon^2)^{3/2}} \right ].
\end{equation}
As detailed in \eqref{eq:Gode}-\eqref{eq:Bode}, the doubly periodic Green's functions can be determined by satisfying the following ODEs in $z$ for $k\neq 0$,

\begin{equation}
    \left( -k^2 + \frac{d^2}{dz^2} \right )\widehat{G}^{DP}_{\phi_{\epsilon}}(z,k) = \widehat{\phi}_{\epsilon}(z),\hspace{.4cm} \left( -k^2 + \frac{d^2}{dz^2} \right ) \widehat{B}^{DP}_{\phi_{\epsilon}}(z,k) = \widehat{G}^{DP}_{\phi_{\epsilon}}(z,k). \label{eq:odeGBapp}
\end{equation}
These ODEs have the particular solutions, 

\begin{align}
    \widehat{G}^{DP}_{\phi_{\epsilon}}(z,k) &= -\frac{\exp \left ( {-k \sqrt{z^2+\epsilon ^2}} \right )} {4 } \left( \frac{\epsilon ^2}{\sqrt{z^2+\epsilon^2}} + \frac{2}{k} \right) \label{eq:GClassic},\\
    \widehat{B}^{DP}_{\phi_{\epsilon}}(z,k) &= \frac{\exp \left ( -k \sqrt{z^2+\epsilon^2} \right )}{4 k^3}\left ( 1+k\sqrt{z^2+\epsilon^2} \right ).
    \label{eq:GBapp}
\end{align}
Unlike in \cite{hoffmann1}, solutions to the homogeneous ODE do not need to be derived to enforce the constraint that $\widehat{G}^{DP}_{\phi_{\epsilon}}, \ \widehat{B}^{DP}_{\phi_{\epsilon}} \to 0$ as $z \to \pm \infty$. Taking the limit of these functions as $\epsilon \to 0$, the singular doubly-periodic Green's functions for $k \neq 0$ as given in \cite{pozrikidis1996} are recovered, 

\begin{equation}
    \widehat{G}_{0}^{DP}(z,k) = -\frac{\exp \left ( -k |z|  \right )} {2 k}, \hspace{.4cm}
    \widehat{B}_{0}^{DP}(z,k) = \frac{\exp \left ( -k |z| \right )}{4 k^3}\left ( 1+k|z| \right ).
    \label{eq:BSingular}
\end{equation}
Taking the derivative of $\widehat{B}^{DP}_{\phi_{\epsilon}}$ in \eqref{eq:GBapp} with respect to $z$ and then evaluating at $z=0$, one can verify that this is always zero, as was claimed in the main text. In the case of the zero wavenumber, $k=0$ ($\alpha=\beta=0$), we can see from the form of the regularized Fourier coefficient matrix of $\widehat{\boldsymbol{\mathcal{S}}}_{\phi_\epsilon}^{DP}$ in \eqref{eq:Scoeff} that we only require the zero wavenumber evaluation of $\widehat{G}_{\phi_{\epsilon}}$ (since the coefficients of all terms with $\widehat{B}_{\phi_{\epsilon}}$include $\alpha$ and/or $\beta$). Solving the ODE for $\widehat{G}_{\phi_{\epsilon}}^{DP}$ in \eqref{eq:odeGBapp} with $k=0$  gives

\begin{equation}
    \widehat{G}^{DP}_{\phi_{\epsilon}}(z,0) = \frac{1}{4}\left ( \frac{2z^2+\epsilon^2}{\sqrt{z^2+\epsilon^2}} \right ).
\end{equation}Notice that $\widehat{G}_{\phi_{\epsilon}}(z;0)$ diverges linearly as $z \to \pm\infty$. In the case of a single regularized force, this would translate into a divergent velocity field. However, if the net force in the domain $R_L \times \mathbb{R}$ is zero, the contributions from the zero wavenumber terms cancel out and the resultant velocity field is convergent \cite{hoffmann1}.

 \section{Doubly-periodic Regularized Stokeslets for Blob Functions with Moment Conditions}\label{app:blobmoment}
\renewcommand{\theequation}{C.\arabic{equation}}
\setcounter{equation}{0}
\noindent \textbf{} \\
The doubly-periodic regularized Stokeslets for the blob functions with moment conditions can be derived in a similar way to the method described in \ref{app:DPSB}. Below, we record $\widehat{\phi}_{\epsilon}^{DP}$ and the associated Green's functions, $\widehat{G}^{DP}_{\phi_{\epsilon}}$ and $\widehat{B}^{DP}_{\phi_{\epsilon}}$ for the other blob functions used in the manuscript. The doubly-periodic Fourier transforms of the blob functions, as well as the solutions of the associated ODEs for the Green's functions, were solved with the aid of Mathematica. \\

\noindent 

\noindent \textbf{Gaussian blob} $\phi_{\epsilon}=\mathbf{\phi}_{\epsilon}^{G,1M}$ from Eq. \eqref{eq:gaussianBlob} \\
The doubly-periodic regularized Stokeslet for this blob function was originally derived in \cite{hoffmann1}. \\ 

\begin{align}
    \widehat{\phi}_\epsilon^{DP}(z,k) &= \frac{\exp \left ( -\frac{k^2 \epsilon ^2} {4}-\frac{z^2}{\epsilon ^2} \right ) \left(k^2 \epsilon ^4-4 z^2+6 \epsilon ^2\right)}{4 \sqrt{\pi } \epsilon ^3} \\ 
    \widehat{G}^{DP}_{\phi_{\epsilon}}(z,k)&= -\frac{\exp \left ( -\frac{k^2 \epsilon ^2}{4}-\frac{z^2}{\epsilon ^2} \right )}{4 \sqrt{\pi}} \nonumber \\ 
    & -\frac{\exp(kz) \text{erfc}\left(\frac{k \epsilon }{2}+\frac{z}{\epsilon }\right)+\exp(-kz) \text{erfc}\left(\frac{k \epsilon }{2}-\frac{z}{\epsilon }\right)}{4 c}, \ \ k \neq 0\\ 
    \widehat{G}^{DP}_{\phi_{\epsilon}}(z,0)&= \frac 1 4 \left [ \frac{\epsilon}{\sqrt \pi} \exp \left (\frac{-z^2}{\epsilon^2} + 2z \text{erf} \left ( \frac z \epsilon \right )\right ) \right ] \\
    \widehat{B}^{DP}_{\phi_{\epsilon}}(z,k) &= \frac{\exp \left ( k z \right ) (1-k z) \text{erfc}\left(\frac{k \epsilon }{2}+\frac{z}{\epsilon }\right)}{8  k^3} \nonumber \\ &+\frac{\exp \left ( -k z \right ) (1+kz) \text{erfc}\left(\frac{k \epsilon }{2}-\frac{z}{\epsilon }\right)}{8  k^3}+\frac{\epsilon  \exp \left (-\frac{k^2 \epsilon ^2}{4}-\frac{z^2}{\epsilon ^2} \right )}{4 \sqrt{\pi} k^2}, \ \ k\neq 0
\end{align}

\noindent \textbf{Gaussian blob} $\phi_{\epsilon}=\mathbf{\phi}_{\epsilon}^{G,3M}$ from Eq. \eqref{eq:gaussianBlob2} \\

\begin{align}
    \phi_{\epsilon}^{DP}(z,k) &= \frac{\exp \left ( -\frac{k^2 \epsilon ^2}{4}-\frac{z^2}{\epsilon ^2} \right )} { 48 \sqrt{\pi } \epsilon ^7}\Bigg [ 48 z^4 \epsilon ^2 \left(k^2 \epsilon ^2+10\right) \nonumber \\ 
    &-12 z^2 \epsilon ^4 \left(k^4 \epsilon ^4+12 k^2 \epsilon ^2+64\right) \nonumber \\
    &+ \epsilon ^6 \left(k^6 \epsilon ^6+6 k^4 \epsilon ^4+48 k^2 \epsilon ^2+192\right)-64 z^6 \Bigg ]  \\
    \widehat{G}^{DP}_{\phi_{\epsilon}}(z,k)&= -\frac{\exp \left ( -\frac{k^2 \epsilon ^2}{4}-\frac{z^2}{\epsilon ^2} \right )}{48 \sqrt{\pi} \epsilon ^3} \Big [ -8 z^2 \epsilon ^2 \left(k^2 \epsilon ^2+6\right) \nonumber \\ 
    &+\epsilon ^4 \left(k^4 \epsilon ^4+4 k^2 \epsilon ^2+24\right)+16 z^4 \Big ] \nonumber \\ 
    &-\frac{\exp(kz) \text{erfc}\left(\frac{k \epsilon }{2}+\frac{z}{\epsilon }\right)+\exp(-kz) \text{erfc}\left(\frac{k \epsilon }{2}-\frac{z}{\epsilon }\right)}{4 c}, \ \ k \neq 0\\
    \widehat{G}^{DP}_{\phi_{\epsilon}}(z,0) &= \frac{3 \sqrt{\pi } z \epsilon ^5 \text{erf}\left(\frac{z}{\epsilon }\right)+e^{-\frac{z^2}{\epsilon ^2}} \left(6 z^2 \epsilon ^4-2 z^4 \epsilon ^2\right)}{6 \sqrt \pi \epsilon ^5} \\
    \widehat{B}^{DP}_{\phi_{\epsilon}}(z,k)&= \frac{1}{8  k^3} \Big [ \exp \left ( k z \right ) (1-k z) \text{erfc}\left(\frac{k \epsilon }{2}+\frac{z}{\epsilon }\right) \nonumber \\
    &+ \exp ( -k z ) (k z+1) \text{erfc}\left(\frac{k \epsilon }{2}-\frac{z}{\epsilon }\right) \Big ] \nonumber \\ 
    &+\frac{\exp \left ( {-\frac{k^2 \epsilon ^2}{4}-\frac{z^2}{\epsilon ^2}} \right )}{48 \sqrt{\pi} k^2} \left(k^2 \epsilon  \left(2 \epsilon ^2-4 z^2\right)+k^4 \epsilon ^5+12 \epsilon \right), \ \ k\neq 0 
\end{align}

\noindent \textbf{Algebraic blob } $\phi_{\epsilon}=\mathbf{\phi}_{\epsilon}^{A,3M}$ from Eq. \eqref{eq:algBlob2} \\

\noindent In the following, let $Z_{\epsilon}=\sqrt{z^2+\epsilon^2}$.
\begin{align}
    \widehat{\phi}^{DP}_{\epsilon}(z,k) &= \frac{ \exp \left ( {-k Z_{\epsilon }} \right ) \epsilon^4}{8  Z_{\epsilon }^{11}} \Bigg [ k Z_{\epsilon } \Bigg ( 3 z^6 \left(3 k^2 \epsilon ^2-2\right) \nonumber \\
    &+z^4 \epsilon ^2 \left(\frac{k^4 \epsilon ^4}{3}-13 k^2 \epsilon ^2+117\right) +2 z^2 \epsilon ^4 \left(\frac{k^4 \epsilon ^4}{3}-9 k^2 \epsilon ^2-84\right)  \nonumber \\
    &+\epsilon ^6 \left(\frac{k^4 \epsilon ^4}{3}+4 k^2 \epsilon ^2+24\right) \Bigg ) -z^6 \left(4 k^4 \epsilon ^4-46 k^2 \epsilon ^2+6\right) \nonumber \\ 
    &-z^4 \epsilon ^2 \left(7 k^4 \epsilon ^4+30 k^2 \epsilon ^2-117\right)-2 z^2 \epsilon ^4 \left(k^4 \epsilon ^4+33 k^2 \epsilon ^2+84\right)  \nonumber \\
    &-2 k^2 z^8+\epsilon ^6 \left(k^4 \epsilon ^4+12 k^2 \epsilon ^2+24\right) \Bigg ] \\
    \widehat{G}^{DP}_{\phi_{\epsilon}}(z,k) &= \frac{\exp \left (-k Z_{\epsilon} \right )}{24 k Z_{\epsilon}^7} \Bigg [ -3 k^2 \epsilon ^4 \left(-2 z^2 \epsilon ^2+z^4+2 \epsilon ^4\right) Z_{\epsilon } \nonumber \\ 
    &-k \epsilon ^2 \Big (z^4 \left(21 \epsilon ^2-4 k^2 \epsilon ^4\right)-2 z^2 \epsilon ^4 \left(k^2 \epsilon ^2-6\right) \nonumber \\
    &+2 \epsilon ^6 \left(k^2 \epsilon ^2+6\right)+6 z^6 \Big)-k^4 \epsilon ^8 Z_{\epsilon }^3-12 Z_{\epsilon }^7 \Bigg], \ \ k\neq 0 \\
    \widehat{G}^{DP}_{\phi_{\epsilon}}(z,0) &= \frac{1}{8 Z_{\epsilon}^7} \left ( 14 z^6 \epsilon ^2+17 z^4 \epsilon ^4+12 z^2 \epsilon ^6+4 z^8 \right )\\
    \widehat{B}^{DP}_{\phi_{\epsilon}}(z,k) &= \frac{\exp \left ( -k Z_\epsilon \right )}{24 k^3 Z_{\epsilon}^3} \left (k^4 \epsilon ^6 Z_{\epsilon }+3 \left(k^2 \epsilon ^2+2\right) Z_{\epsilon }^3+k^3 \epsilon ^6+6 k Z_{\epsilon }^4 \right ), \ \ k\neq 0
\end{align}

 \section{Doubly-periodic Regularized Stokeslets with IFFTs}\label{app:ffts}
\renewcommand{\theequation}{D.\arabic{equation}}
\setcounter{equation}{0}
To simulate the elastic surface in the doubly-periodic domain $R_L \times \mathbb{R}$, we use the pseudo-spectral method described in \cite{hoffmann1} to evaluate the velocity at every time step. This requires using Inverse Fast Fourier Transforms (IFFTs) to go from the representation of the velocity field in Fourier space back to physical space. 
Let \[ \vec X(q_j,s_k,t)=(X^1(q_j,s_k,t), X^2(q_j,s_k,t), X^3(q_j,s_k,t))\]represent the discrete elastic surface, with $j,k \in \Zmodn$ and $(q_j,s_k)=(jh,kh)$ in the reference configuration and $h=L/N$. Suppose we wish to evaluate the velocity on a uniform grid for a fixed value of $z$. Let $(x_m,y_n,z)=(mh,nh,z)$ with $m,n \in \Zmodn$. At every instant in the simulation, we follow the procedure in Alg. \ref{alg:fft}.

\renewcommand{\algorithmiccomment}[1]{// #1}
\begin{algorithm*}[t!]
  \caption{Doubly Periodic Stokeslets with IFFTs}\label{alg:fft}
  \begin{algorithmic}
   \State $\whvec u_{\alpha,\beta}(z) \gets 0$ for $\alpha,\beta \in \Fmodn \times \Fmodn$
   \State{Evaluate $\vec F_{j,k}$ using finite differences based on elastic model being used, for $(j,k) \in \Zmodn \times \Zmodn$}. 
   
   \For{$(\alpha, \beta) \in \Fmodn \times \Fmodn $} 

   \State \begin{align*}
         \whvec u_{\alpha,\beta}(z;t) \gets \whvec u_{\alpha,\beta}(z;t) + \frac{1}{L^2}&\sum_{j,k}\frac{1}{8\pi \mu} \Sdp(z-X^3(q_j,s_k;t); \alpha,\beta) \cdot \vec F_{j,k} \ h^2 \cdot \\ &\exp \left ( - \frac{2\pi i}{L} \left( X^1(q_j,s_k,t) \alpha + X^2(q_j,s_k,t) \beta \right )\right ). 
    \end{align*} 
    \Comment{The exponential in the sum specifies the location of the force in the periodic directions.}

   \EndFor
    \State    $\vec u(x_m,y_n,z; t) \gets$  IFFT$(\whvec u_{\alpha,\beta}(z;t))$ \\
   
  \end{algorithmic}
\end{algorithm*}

    To get the velocity at the surface points, we evaluated the velocity at several $z$-levels and then interpolated from the $z$-levels to the surface points. For better accuracy between nodes on the surface, we used trigonometric interpolation through the IFFTs on each $z$-level. This allowed us to evaluate the velocity at more points for each $z$-level which reduced errors due to interpolation.

    As explained in \cite{hoffmann1}, an Ewald splitting that splits the velocity into a local, fast-decaying part, and a long-ranged, smooth part, may be important for accuracy when $\epsilon$ shrinks below a certain value. This is due to the fact that when $\epsilon$ is too small relative to the grid, the blob function (and consequently the associated force) is not captured well on the grid. For the Gaussian blobs $\phi^{G,1M}_{\epsilon}$ and $\phi^{G,3M}_{\epsilon}$, we used the Ewald splitting when $\epsilon/h \leq 4$ for $\phi^{G,1M}_{\epsilon}$, and when $\epsilon/h \leq 8$ for $\phi^{G,3M}_{\epsilon}$. We set the splitting parameter $\xi = 6h$ and set the cutoff parameter to be $1.75$ times the length of $L$.  For more details on these parameter choices and Ewald splitting for doubly periodic regularized Stokeslets, we refer the reader to \cite{hoffmann1}. 

    In order to implement the Ewald splitting, one needs the regularized Green's functions for the blob functions $\phi_{\epsilon}^{G,1M}$ (Eq. \eqref{eq:gaussianBlob}) and $\phi_{\epsilon}^{G,3M}$ (Eq. \eqref{eq:gaussianBlob2}) in an unbounded fluid. These are as follows: 

    \begin{align}
        \phi_{\epsilon}^{G,1M}&: \nonumber \\
        G(r) &= -\frac{\text{erf}\left(\frac{r}{\epsilon }\right)}{4 \pi  r}-\frac{e^{-\frac{r^2}{\epsilon ^2}}}{4 \pi ^{3/2} \epsilon }, \\
       B(r) &= -\frac{r \text{erf}\left(\frac{r}{\epsilon }\right)}{8 \pi }-\frac{\epsilon  e^{-\frac{r^2}{\epsilon ^2}}}{8 \pi ^{3/2}}, \\[2em]
       \phi_{\epsilon}^{G,3M}&: \nonumber \\
        G(r)&= -\frac{e^{-\frac{r^2}{\epsilon ^2}} \left(3 \sqrt{\pi } \epsilon ^5 e^{\frac{r^2}{\epsilon ^2}} \text{erf}\left(\frac{r}{\epsilon }\right)-20 r^3 \epsilon ^2+4 r^5+18 r \epsilon ^4\right)}{12 \pi ^{3/2} r \epsilon ^5}, \\
        B(r)&= \frac{r \left(-3 \sqrt{\pi } \text{erf}\left(\frac{r}{\epsilon }\right)-\frac{2 r e^{-\frac{r^2}{\epsilon ^2}}}{\epsilon }\right)}{24 \pi ^{3/2}}.
    \end{align}
    
    Unfortunately, the Ewald splitting could not be used for the algebraic blobs. This is because the strategy employed for the Gaussian blobs does not result in a fast-decaying, local part for the algebraic blobs. Despite this, we gained confidence from the pure IFFT method described in Algorithm \ref{alg:fft} by the convergence rate test described in \cite{hoffmann1}: we placed random forces with zero net force in the doubly periodic domain $R_L \times \mathbb{R}$ and then evaluated the doubly periodic velocity field. This can be compared to the velocity field generated by placing $M^2$ copies of these forces in the periodic lattice. For example, given a force $\vec f_0$ located at $(x_0,y_0,z_0)$ in the original domain, the velocity at a point $(x,y,z)$ is evaluated as 

    \begin{align*}
    \vec u(x,y,z) &= \sum_{j,k=-M/2}^{M/2-1} \frac{1}{8\pi \mu}\Stokes(\whvec r_{j,k}) \cdot \vec f_0, \\ 
    \vec r_{j,k}&=(x-x_0-jL, y-y_0-kL, z-z_0).
    \end{align*}Performing this for multiple $M$, the empirical convergence rate of this velocity field with the doubly periodic velocity field is evaluated and compared to the theoretical convergence rate, which is $\mathcal{O}(1/M)$. We were able to confirm this expected convergence rate for the two larger regularization parameters used for the algebraic blobs. The smallest regularization parameters ($0.03425$ for $\phi_{\epsilon}^A$ and $0.09325$ for $\phi_{\epsilon}^{A,3M})$ had slower convergence rates but for completeness, we have reported their numerical results in Table \ref{tab:algebraictables}.

 \section{Supplementary Information}
\label{app:videos}
\renewcommand{\theequation}{E.\arabic{equation}}
\setcounter{equation}{0} 
In all three videos, the surface is discretized with $N=32$ points so that the spatial discretization is $h=1/32$. We use the blob function $\phi_{\epsilon}^A$ with $\epsilon \approx 2.2h$. The surface has tensile rigidity $\sigma_q=\sigma_s=75$ and bending rigidity $\kappa_B=0.075$.
\begin{enumerate}
    \item Video 1 (\textit{stableDP.avi}): Stable, doubly periodic surface using time step $dt = 3.7e-3$
    \item Video 2 (\textit{unstableDP.avi}): Unstable, doubly periodic surface using time step $dt = 3.9e-3$.
    \item Video 3 (\textit{finiteAperiodic.avi}): Stable, finite surface in unbounded fluid using time step $dt=7.5e-4$.
\end{enumerate} 
\newpage
\bibliographystyle{elsarticle-num} 
\bibliography{ReferencesV2}

\begin{thebibliography}{10}
\expandafter\ifx\csname url\endcsname\relax
  \def\url#1{\texttt{#1}}\fi
\expandafter\ifx\csname urlprefix\endcsname\relax\def\urlprefix{URL }\fi
\expandafter\ifx\csname href\endcsname\relax
  \def\href#1#2{#2} \def\path#1{#1}\fi

\bibitem{Griffith20}
B.~Griffith, N.~Patankar, Immersed methods for fluid–structure interaction,
  Annu. Review Fluid Mech. 52 (2020) 421--448.
\newblock \href
  {https://doi.org/https://doi.org/10.1146/annurev-fluid-010719-060228}
  {\path{doi:https://doi.org/10.1146/annurev-fluid-010719-060228}}.

\bibitem{gaffney11}
E.~Gaffney, H.~Gad\^{e}lha, D.~Smith, J.~Blake, J.~Kirkman-Brown, Mammalian
  sperm motility: Observation and theory, Annu. Review Fluid Mech. 43~(1)
  (2011) 501--528.
\newblock \href {https://doi.org/10.1146/annurev-fluid-121108-145442}
  {\path{doi:10.1146/annurev-fluid-121108-145442}}.

\bibitem{Lauga16}
E.~Lauga, Bacterial hydrodynamics, Annu. Review Fluid Mech. 48 (2016) 105--130.
\newblock \href
  {https://doi.org/https://doi.org/10.1146/annurev-fluid-122414-034606}
  {\path{doi:https://doi.org/10.1146/annurev-fluid-122414-034606}}.

\bibitem{Gilpin20}
W.~Gilpin, M.~Bull, M.~Prakash, The multiscale physics of cilia and flagella,
  Nat. Rev. Phys. 2 (2020) 74--88.
\newblock \href {https://doi.org/https://doi.org/10.1038/s42254-019-0129-0}
  {\path{doi:https://doi.org/10.1038/s42254-019-0129-0}}.

\bibitem{Salathe07}
M.~Salathe, Regulation of mammalian ciliary beating, Annu. Rev. Phys. 69 (2007)
  401--422.
\newblock \href
  {https://doi.org/https://doi.org/10.1146/annurev.physiol.69.040705.141253}
  {\path{doi:https://doi.org/10.1146/annurev.physiol.69.040705.141253}}.

\bibitem{Forth17}
S.~Forth, T.~Kapoor, The mechanics of microtubule networks in cell division, J.
  Cell Biol. 216 (2017) 1525--1531.
\newblock \href {https://doi.org/https://doi.org/10.1083/jcb.201612064}
  {\path{doi:https://doi.org/10.1083/jcb.201612064}}.

\bibitem{Kent17}
I.~Kent, T.~Lele, Microtubule-based force generation, WIREs Nanomed.
  Nanobiotechnol. 9 (2017) e1428.
\newblock \href {https://doi.org/https://doi.org/10.1002/wnan.1428}
  {\path{doi:https://doi.org/10.1002/wnan.1428}}.

\bibitem{Freund14}
J.~Freund, Numerical simulation of flowing blood cells, Annu. Review Fluid
  Mech. 46 (2014) 67--95.
\newblock \href
  {https://doi.org/https://doi.org/10.1146/annurev-fluid-010313-141349}
  {\path{doi:https://doi.org/10.1146/annurev-fluid-010313-141349}}.

\bibitem{Biesel16}
D.~Barthès-Biesel, Motion and deformation of elastic capsules and vesicles in
  flow, Annu. Review Fluid Mech. 48 (2016) 25--52.
\newblock \href
  {https://doi.org/https://doi.org/10.1146/annurev-fluid-122414-034345}
  {\path{doi:https://doi.org/10.1146/annurev-fluid-122414-034345}}.

\bibitem{Nelson23}
B.~Nelson, S.~Pane, Delivering drugs with microrobots, Science 382~(6675)
  (2023) 1120--1122.
\newblock \href {https://doi.org/https://doi.org/10.1126/science.adh3073}
  {\path{doi:https://doi.org/10.1126/science.adh3073}}.

\bibitem{Espina23}
J.~A. Espina, M.~H. Cordeiro, M.~Milivojevic, I.~Pajić-Lijaković, E.~H.
  Barriga, Response of cells and tissues to shear stress, J. Cell Sci. 136~(18)
  (2023).
\newblock \href {https://doi.org/https://doi.org/110.1242/jcs.260985}
  {\path{doi:https://doi.org/110.1242/jcs.260985}}.

\bibitem{Rallabandi24}
B.~Rallabandi, Fluid-elastic interactions near contact at low reynolds number,
  Annu. Review Fluid Mech. 56 (2024) 491--519.
\newblock \href
  {https://doi.org/https://doi.org/10.1146/annurev-fluid-120720-024426}
  {\path{doi:https://doi.org/10.1146/annurev-fluid-120720-024426}}.

\bibitem{pozr1992}
C.~Pozrikidis, Boundary Integral and Singularity Methods for Linearized Viscous
  Flow, Cambridge Texts in Applied Mathematics, Cambridge University Press,
  1992.
\newblock \href {https://doi.org/https://doi.org.10.1017/CBO9780511624124}
  {\path{doi:https://doi.org.10.1017/CBO9780511624124}}.

\bibitem{veerapaneni09}
S.~K. Veerapaneni, D.~Gueyffier, D.~Zorin, G.~Biros, A boundary integral method
  for simulating the dynamics of inextensible vesicles suspended in a viscous
  fluid in 2d, Journal of Computational Physics 228~(7) (2009) 2334--2353.

\bibitem{peskin}
C.~S. Peskin, The immersed boundary method, Acta Numer. 11 (2002) 479--517.
\newblock \href {https://doi.org/https://doi.org/10.1017/S0962492902000077}
  {\path{doi:https://doi.org/10.1017/S0962492902000077}}.

\bibitem{cortez2001}
R.~Cortez, The method of regularized {S}tokeslets, SIAM J. Sci. Comput. 23~(4)
  (2001) 1204--1225.
\newblock \href {https://doi.org/https://doi.org/10.1137/S106482750038146X}
  {\path{doi:https://doi.org/10.1137/S106482750038146X}}.

\bibitem{cortez2005}
R.~Cortez, L.~Fauci, A.~Medovikov, The method of regularized {S}tokeslets in
  three dimensions: Analysis, validation, and application to helical swimming,
  Physics of Fluids 17 (2005) 031504.
\newblock \href {https://doi.org/https://doi.org/10.1063/1.1830486}
  {\path{doi:https://doi.org/10.1063/1.1830486}}.

\bibitem{smith18}
D.~J. Smith,
  \href{https://www.sciencedirect.com/science/article/pii/S0021999117308847}{A
  nearest-neighbour discretisation of the regularized {S}tokeslet boundary
  integral equation}, J. Comp. Phys. 358 (2018) 88--102.
\newblock \href {https://doi.org/https://doi.org/10.1016/j.jcp.2017.12.008}
  {\path{doi:https://doi.org/10.1016/j.jcp.2017.12.008}}.
\newline\urlprefix\url{https://www.sciencedirect.com/science/article/pii/S0021999117308847}

\bibitem{smithDouble}
D.~J. Smith, M.~T. Gallagher, R.~Schuech, T.~D. Montenegro-Johnson, The role of
  the double-layer potential in regularised {S}tokeslet models of
  self-propulsion, Fluids 6~(11) (2021).
\newblock \href {https://doi.org/https://doi.org/10.3390/fluids6110411}
  {\path{doi:https://doi.org/10.3390/fluids6110411}}.

\bibitem{Needleman19}
D.~Needleman, M.~Shelley, The stormy fluid dynamics of the living cell, Physics
  Today 72~(9) (2019) 32--38.
\newblock \href {https://doi.org/https://doi.org/10.1063/PT.3.4292}
  {\path{doi:https://doi.org/10.1063/PT.3.4292}}.

\bibitem{tornberg2004}
A.~Tornberg, M.~Shelley, Simulating the dynamics and interactions of flexible
  fibers in {S}tokes flows, J. Comp. Phys. 196~(1) (2004) 8--40.
\newblock \href {https://doi.org/https://doi.org/10.1016/j.jcp.2003.10.017}
  {\path{doi:https://doi.org/10.1016/j.jcp.2003.10.017}}.

\bibitem{nazockdast2017}
E.~Nazockdast, A.~Rahimian, D.~Zorin, M.~Shelley, A fast platform for
  simulating semi-flexible fiber suspensions applied to cell mechanics, J.
  Comp. Phys. 329 (2017) 173--209.
\newblock \href {https://doi.org/https://doi.org/10.1016/j.jcp.2016.10.026}
  {\path{doi:https://doi.org/10.1016/j.jcp.2016.10.026}}.

\bibitem{huapeskin}
M.~Hua, C.~S. Peskin, An analysis of the numerical stability of the immersed
  boundary method, J. Comp. Phys. 467 (2022) 111435.
\newblock \href {https://doi.org/https://doi.org/10.1016/j.jcp.2022.111435,}
  {\path{doi:https://doi.org/10.1016/j.jcp.2022.111435,}}.

\bibitem{Gong08}
Z.~Gong, H.~Huang, C.~Lu, Stability analysis of the immersed boundary method
  for a two-dimensional membrane with bending rigidity, Commun. Comp. Phys. 3
  (2008) 704--723.

\bibitem{stockie99}
J.~M. Stockie, B.~R. Wetton, Analysis of stiffness in the immersed boundary
  method and implications for time-stepping schemes, J. Comp. Phys. 154~(1)
  (1999) 41--64.
\newblock \href {https://doi.org/https://doi.org/10.1006/jcph.1999.6297}
  {\path{doi:https://doi.org/10.1006/jcph.1999.6297}}.

\bibitem{stockie95}
J.~Stockie, B.~Wetton, Stability analysis for the immersed fiber problem, SIAM
  J. Appl. Math. 55 (1995) 1577--1591.
\newblock \href {https://doi.org/https://doi.org/10.1137/S0036139994267018}
  {\path{doi:https://doi.org/10.1137/S0036139994267018}}.

\bibitem{Guy15}
R.~Guy, B.~Philip, B.~Griffith, Geometric multigrid for an implicit-time
  immersed boundary method, Adv. Comput. Math. 41 (2015) 635--662.
\newblock \href {https://doi.org/https://doi.org/10.1007/s10444-014-9380-1}
  {\path{doi:https://doi.org/10.1007/s10444-014-9380-1}}.

\bibitem{ainley}
J.~Ainley, S.~Durkin, R.~Embid, P.~Boindala, R.~Cortez, The method of images
  for regularized stokeslets, J. Comp. Phys. 227~(9) (2008) 4600--4616.
\newblock \href {https://doi.org/https://doi.org/10.1016/j.jcp.2008.01.032}
  {\path{doi:https://doi.org/10.1016/j.jcp.2008.01.032}}.

\bibitem{hoffmann2}
R.~Cortez, F.~Hoffmann, A fast numerical method for computing doubly-periodic
  regularized {S}tokes flow in 3{D}, J. Comp. Phys. 258 (2014) 1–14.
\newblock \href {https://doi.org/10.1016/j.jcp.2013.10.032}
  {\path{doi:10.1016/j.jcp.2013.10.032}}.

\bibitem{cortezvarela}
R.~Cortez, D.~Varela, A general system of images for regularized stokeslets and
  other elements near a plane wall, J. Comp. Phys. 285 (2015) 41--54.
\newblock \href {https://doi.org/https://doi.org/10.1016/j.jcp.2015.01.019}
  {\path{doi:https://doi.org/10.1016/j.jcp.2015.01.019}}.

\bibitem{hoffmann1}
F.~Hoffmann, R.~Cortez, Numerical computation of doubly-periodic stokes flow
  bounded by a plane with applications to nodal cilia, Commun. Comput. Phys.
  22~(3) (2017) 620–642.
\newblock \href {https://doi.org/10.4208/cicp.OA-2016-0151}
  {\path{doi:10.4208/cicp.OA-2016-0151}}.

\bibitem{LBCL2013}
K.~Leiderman, E.~Bouzarth, R.~Cortez, A.~Layton, A regularization method for
  the numerical solution of periodic {S}tokes flow, J. Comp. Phys. 236~(236)
  (2013) 187--202.
\newblock \href {https://doi.org/https://doi.org/10.1016/j.jcp.2012.09.035}
  {\path{doi:https://doi.org/10.1016/j.jcp.2012.09.035}}.

\bibitem{mannan}
F.~Mannan, R.~Cortez, An explicit formula for two-dimensional singly-periodic
  regularized stokeslets flow bounded by a plane wall, Commun. Comput. Phys.
  23~(1) (2018) 142--167.

\bibitem{tripBrinkman}
H.-N. Nguyen, S.~Olson, K.~Leiderman, A fast method to compute triply-periodic
  brinkman flows, Comput. Fluids 133 (2016) 55--67.
\newblock \href
  {https://doi.org/https://doi.org/10.1016/j.compfluid.2016.04.007}
  {\path{doi:https://doi.org/10.1016/j.compfluid.2016.04.007}}.

\bibitem{Zheng23}
P.~Zheng, D.~Apsley, S.~Zhong, J.~Sznitman, A.~Smits, Image systems for
  regularised {S}tokeslets at walls and free surfaces, Eur. Mech. B Fluids 97
  (2023) 112--127.
\newblock \href
  {https://doi.org/https://doi.org/10.1016/j.euromechflu.2022.09.005}
  {\path{doi:https://doi.org/10.1016/j.euromechflu.2022.09.005}}.

\bibitem{bouzarth2010}
E.~L. Bouzarth, M.~L. Minion, A multirate time integrator for regularized
  {S}tokeslets, J. Comp. Phys. 229~(11) (2010) 4208--4224.
\newblock \href {https://doi.org/https://doi.org/10.1016/j.jcp.2010.02.006}
  {\path{doi:https://doi.org/10.1016/j.jcp.2010.02.006}}.

\bibitem{Liu22}
W.~Liu, M.~Rostami, Parallel-in-time simulation of biofluids, J. Comp. Phys.
  464 (2022) 111366.
\newblock \href {https://doi.org/https://doi.org/10.1016/j.jcp.2022.111366}
  {\path{doi:https://doi.org/10.1016/j.jcp.2022.111366}}.

\bibitem{Liu23}
W.~Liu, M.~Rostami, A multigrid method for kernel functions acting on
  interacting structures with applications to biofluids, J. Comp. Phys. 494
  (2023) 112506.
\newblock \href {https://doi.org/https://doi.org/10.1016/j.jcp.2023.112506}
  {\path{doi:https://doi.org/10.1016/j.jcp.2023.112506}}.

\bibitem{jabbarzadeh2020}
M.~Jabbarzadeh, H.~C. Fu, A numerical method for inextensible elastic filaments
  in viscous fluids, J. Comp. Phys. 418 (2020) 109643.
\newblock \href {https://doi.org/https://doi.org/10.1016/j.jcp.2020.109643}
  {\path{doi:https://doi.org/10.1016/j.jcp.2020.109643}}.

\bibitem{nguyen2014}
H.-N. Nguyen, R.~Cortez, Reduction of the regularization error of the method of
  regularized {S}tokeslets for a rigid object immersed in a three-dimensional
  {S}tokes flow, Commun. Comput. Phys. 15~(1) (2014) 126--152.
\newblock \href {https://doi.org/https://doi.org/10.4208/cicp.021112.290413a}
  {\path{doi:https://doi.org/10.4208/cicp.021112.290413a}}.

\bibitem{zhao2019}
B.~Zhao, E.~Lauga, L.~Koens, Method of regularized {s}tokeslets: Flow analysis
  and improvement of convergence, Phys. Rev. Fluids 4~(8) (2019) 084104.
\newblock \href
  {https://doi.org/https://doi.org/10.1103/PhysRevFluids.4.084104}
  {\path{doi:https://doi.org/10.1103/PhysRevFluids.4.084104}}.

\bibitem{Chisholm22}
N.~G. Chisholm, S.~D. Olson, A framework for generating radial and
  surface-oriented regularized {S}tokeslets, Fluids 7~(11) (2022).
\newblock \href {https://doi.org/https://doi.org/10.3390/fluids7110351}
  {\path{doi:https://doi.org/10.3390/fluids7110351}}.

\bibitem{nguyen2025}
H.~Nguyen, A.~Gibbs, F.~Healy, O.~Shindell, R.~Cortez, K.~M. Brown, J.~McCoy,
  B.~Rodenborn, Using theory and experiments of spheres moving near boundaries
  to optimize the method of images for regularized stokeslets, Physical Review
  Fluids 10~(3) (2025) 033101.

\bibitem{bgil}
A.~Barrero-Gil, Weakening accuracy dependence with the regularization parameter
  in the method of regularized {S}tokeslets, J. Comput. Appl. Math. 237~(1)
  (2013) 672--679.
\newblock \href {https://doi.org/https://doi.org/10.1016/j.cam.2012.08.014}
  {\path{doi:https://doi.org/10.1016/j.cam.2012.08.014}}.

\bibitem{cortez18}
R.~Cortez, Regularized {S}tokeslet segments, J. Comp. Phys. 375 (2018)
  783--796.
\newblock \href {https://doi.org/https://doi.org/10.1016/j.jcp.2018.08.055}
  {\path{doi:https://doi.org/10.1016/j.jcp.2018.08.055}}.

\bibitem{ferranti2024}
D.~Ferranti, R.~Cortez, Regularized {S}tokeslet surfaces, J. Comp. Phys. 508
  (2024) 113004.
\newblock \href {https://doi.org/https://doi.org/10.1016/j.jcp.2024.113004}
  {\path{doi:https://doi.org/10.1016/j.jcp.2024.113004}}.

\bibitem{Boffi07}
D.~Boffi, L.~Gastaldi, L.~Heltai, Numerical stability of the finite element
  immersed boundary method, Math. Models Methods Appl. Sci. 17~(10) (2007) 1479
  – 1505.
\newblock \href {https://doi.org/htpps://doi.org/10.1142/S0218202507002352}
  {\path{doi:htpps://doi.org/10.1142/S0218202507002352}}.

\bibitem{Heltai08}
L.~Heltai, On the stability of the finite element immersed boundary method,
  Comput. Struct. 86~(7) (2008) 598--617.
\newblock \href
  {https://doi.org/https://doi.org/10.1016/j.compstruc.2007.08.008}
  {\path{doi:https://doi.org/10.1016/j.compstruc.2007.08.008}}.

\bibitem{fauci1988}
L.~J. Fauci, C.~S. Peskin, A computational model of aquatic animal locomotion,
  J. Comp. Phys. 77~(1) (1988) 85--108.
\newblock \href {https://doi.org/https://doi.org/10.1016/0021-9991(88)90158-1}
  {\path{doi:https://doi.org/10.1016/0021-9991(88)90158-1}}.

\bibitem{Landau1970}
L.~Landau, E.~Lifshitz, Theory of Elasticity, 2nd Edition, Pergamon Press,
  1970.

\bibitem{pozr2011}
C.~Pozrikidis, Introduction to theoretical and computational fluid dynamics,
  2nd Edition, Oxford University Press, New York, NY, 2011.

\bibitem{olson2011}
S.~D. Olson, S.~S. Suarez, L.~J. Fauci, Coupling biochemistry and hydrodynamics
  captures hyperactivated sperm motility in a simple flagellar model, J. Theor.
  Biol. 283~(1) (2011) 203--216.
\newblock \href {https://doi.org/https://doi.org/10.1016/j.jtbi.2011.05.036}
  {\path{doi:https://doi.org/10.1016/j.jtbi.2011.05.036}}.

\bibitem{Yu2022}
Y.~Yu, M.~D. Graham, Wrinkling and multiplicity in the dynamics of deformable
  sheets in uniaxial extensional flow, Phys. Rev. Fluids 7 (2022) 023601.
\newblock \href
  {https://doi.org/https://doi.org/10.1103/PhysRevFluids.7.023601}
  {\path{doi:https://doi.org/10.1103/PhysRevFluids.7.023601}}.

\bibitem{beale2001}
J.~Beale, A convergent boundary integral method for three-dimensional water
  waves, Math. Comput. 70~(235) (2001) 977--1029.
\newblock \href {https://doi.org/https://doi.org/10.1090/S0025-5718-00-01218-7}
  {\path{doi:https://doi.org/10.1090/S0025-5718-00-01218-7}}.

\bibitem{ascher2008}
U.~M. Ascher, Numerical Methods for Evolutionary Differential Equations,
  Society for Industrial and Applied Mathematics, USA, 2008.

\bibitem{pozrikidis1996}
C.~Pozrikidis, Computation of periodic {G}reen's functions of {S}tokes flow, J.
  Eng. Math. 30~(1) (1996) 79--96.
\newblock \href {https://doi.org/https://doi.org/10.1007/BF00118824}
  {\path{doi:https://doi.org/10.1007/BF00118824}}.

\end{thebibliography}
\end{document}